\documentclass[12pt]{article}

\usepackage{epsfig}
\usepackage{amsmath}
\usepackage{amsfonts}
\usepackage{amssymb}
\usepackage{hyperref}

\newcommand{\be}{\begin{eqnarray}}
\newcommand{\ee}{\end{eqnarray}}
\newcommand{\non}{\nonumber\\}
\newcommand{\ave}[1]{\left\langle #1 \right\rangle}

\newcommand{\absol}[1]{\left| #1 \right|}

\newcommand{\rb}{\right)}
\newcommand{\lb}{\left(}

\newcommand{\QGP}{Quark Gluon Plasma }

\title{Hadronic Fluctuations and Correlations}

\author{Volker Koch \\
Nuclear Science Division\\
Lawrence Berkeley National Laboratory\\
Berkeley, CA 94720, USA
}

\begin{document}

\maketitle

\begin{abstract}
We will provide a review of some of the physics which can be addressed by studying fluctuations and correlations in heavy ion collisions. We will discuss Lattice QCD results on fluctuations and correlations and will put them into context with observables which have been measured in heavy-ion collisions. Special attention will be given to the QCD critical point and the first order co-existence region, and we will discuss how the measurement of fluctuations and correlations can help in an experimental search for non-trivial structures in the QCD phase diagram.
\end{abstract}

\section{Introduction}
\label{sec:introduction}
Fluctuations and correlations are important characteristics of any physical system. They  provide essential information about
the effective degrees of freedom and their possible quasi-particle nature. In addition,  the susceptibilities, 
which characterize the correlations and fluctuations,  determine the response of the system to small external forces. 

In general, one can distinguish between several classes of
fluctuations. On the most fundamental level there are quantum fluctuations,
which arise if the specific observable  does not commute with the
Hamiltonian of the system under consideration. These fluctuations probably
play less a role for the physics of heavy-ion collisions. Second, there are
``dynamical'' fluctuations and correlations reflecting the underlying dynamics of the system.  Examples are density fluctuations, which are controlled by the compressibility of the system. Finally, there are ``trivial'' fluctuations induced by the measurement process itself, such as finite number statistics, etc. These need to be understood, controlled and subtracted in order to access the dynamical fluctuations which tell as about the properties of the system.

A prominent example where the measurement of correlations has lead to a scientific breakthrough are the fluctuations of the cosmic microwave background radiation \cite{Smoot:2007zz}, first carried out by the COBE satelite \cite{Smoot:1992td} and later refined by WMAP \cite{Spergel:2006hy}. In the case of the cosmic microwave background, the fluctuations are on the level of $10^{-4}$ with respect to the thermal distribution. In addition, a large dipole correlation, due to the motion of the earth through the heatbath of the microwave background, has been seen and needed to be subtracted before the more interesting correlations due to the big bang, could be revealed.

In the cosmic microwave background one observes spatial correlations. 
In contrast, in heavy-ion collisions we are restricted to correlations in momenta and the connection to spatial correlations is, in general, non-trivial. Nonetheless, in the case of heavy-ion collisions  we have a quite  similar situation. To first order we observe thermal spectra and radial flow. Next, we see a large quadrupole correlation due to elliptic flow. The interesting question then remains, if we will be able to identify smaller correlations due to a possible phase transition and the QCD critical point after subtracting the two dominant backgrounds, thermal emission and elliptic flow.  

\begin{figure}[th]
\begin{center}
    \includegraphics[width=0.5\textwidth]{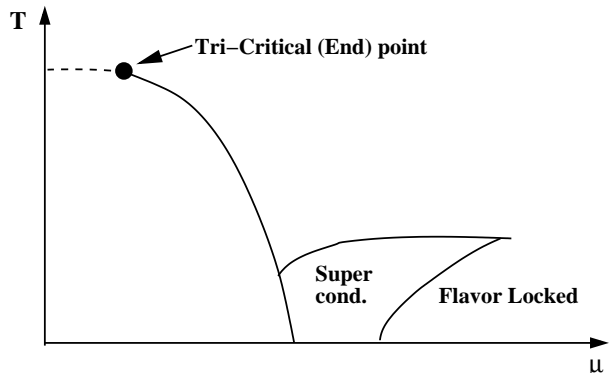}
\end{center}
\caption{Schematic QCD phase diagram.}
\label{fig:phase_diagram}
\end{figure}

Fluctuations are closely related to phase transitions. The well known
phenomenon of critical opalescence is a result of fluctuations at all
length scales due to a second order phase transition. First order transitions,
on the other hand, give rise to bubble formation, i.e. density fluctuations at
the extreme. Considering the richness of the QCD phase-diagram, as sketched in
Fig.\ref{fig:phase_diagram},
the study of fluctuations in heavy-ion physics should lead to a rich set of
phenomena and is an essential tool for the experimental exploration of the QCD phase diagram.

The most efficient way to address fluctuations of a system created in a heavy-ion
collision is via the study of event-by-event (E-by-E) fluctuations, where
a given observable is measured on an event-by-event basis and the fluctuations
are studied over the ensemble  of the events. Equivalently, one may also study  appropriate multi-particle correlations within the same region of acceptance \cite{Bialas:1999tv}. 

In the framework of statistical physics, which appears to describe the bulk
properties of heavy-ion collisions up to RHIC energies \cite{Braun-Munzinger:2003zd}, fluctuations
measure the susceptibilities of the system. These susceptibilities also
determine the response of the system to external forces. For example, by
measuring fluctuations of the net electric charge 
in a given rapidity interval, one
obtains information on how this (sub)system would respond to applying an
external (static) electric field. In
other words, by measuring fluctuations one gains access to the same
fundamental properties of the system as ``table top'' experiments dealing with
macroscopic probes. In the later case, of course, fluctuation measurements
would be impossible.

In addition, the study of fluctuations may reveal information beyond its
thermodynamic properties. If the system expands, fluctuations may be frozen in
early and thus tell us about the properties of the system prior to its
thermal freeze out. A  well known example is the fluctuations in the
cosmic microwave background radiation. Similarly, in the context of heavy-ion collisions, one finds that fluctuations of conserved charges of a subsystem exhibit relaxation times which are much longer than the typical equilibration time of the system \cite{Jeon:2000wg,Shuryak:2000pd}.

While the  field of event-by-event fluctuations is comparatively new to heavy-ion physics, quite a number of measurements have been carried out. Most are concerned with multiplicity fluctuations, fluctuations of particle ratios, transverse momentum, and net-electric charge. In this review, we will give an overview of the underlying concepts motivating these measurements. While discussing these fluctuations we will keep in close contact to results from Lattice QCD calculations, where fluctuations and correlations can be calculated for a thermal system. We will also discuss how the careful study of fluctuations may help to unravel the QCD phase diagram. In particular, the question about the existence of a critical point and its experimental identification  will have to involve a detailed analysis of the fluctuations and correlations of the system. 

This review is organized as follows. In the following section we will introduce the basic definitions and concepts in the framework of a thermal system. Next we will discuss fluctuations and correlations of conserved charges. These have a direct mapping on Lattice QCD and have the advantage that their relaxation time is rather long, and thus should be less affected by hadronic re-interactions. We will then turn to the discussion of observables which have actually been measured and try to put them into context with recent lattice calculations. The last section will be devoted to a discussion on how the study of fluctuations may help in the search for the QCD critical point and the first order phase co-existence region.

Finally, let us emphasize that this article is not intended to be a comprehensive review of all aspects concerning fluctuations and correlation in heavy-ion physics. Our main emphasis is on the properties of the bulk matter created in these collisions, which we discuss based on a subset of phenomena and observables. For example, multiplicity fluctuations will be touched upon only briefly. Also, the interesting correlation structure observed among high momentum particles will not be discussed.

\section{Fluctuations and Correlations in a thermal system and from Lattice QCD}
\label{sec:thermal}

The concepts of fluctuations and correlations have a well defined physical interpretation for a system in thermal equilibrium. In this case fluctuations and correlations are related to the second cumulants of the partition function. These cumulants, or susceptibilities, can also be expressed in terms of integrals of equal-time correlation functions, which in turn characterize the (space-like) static responses of the system. In the case of heavy-ion collisions other, non-statistical, effects may contribute to the measured correlations. For example, the dynamical evolution of the system may be too fast for long range correlations to build up. These effects need to be understood well in order to extract the matter properties from these experiments and we will briefly discuss some of the issues in section \ref{sec:observable}. 

As already mentioned, the study of fluctuations is essential for the characterization of phase transitions. In case of a second order phase transition, the fluctuations of the order parameter diverge with a critical exponent specific to the universality class of the transition \cite{Landau_Stat}. Furthermore, the system size dependences of the fluctuations can be used to distinguish between cross-over transitions and first or second order phase transitions using finite size scaling arguments. 

A strongly interacting system in thermal equilibrium can be studied in the framework of Lattice QCD (LQCD) (see e.g. \cite{Karsch:2003jg} for a review).
Here we will just use the results from LQCD as input to discuss the relevant physics\footnote{We note that at present most lattice calculations are using rather high quark masses, corresponding to pion masses of the order of several hundred MeV. However, recently first results with physical strange quark mass and light quark masses corresponding to a pion mass of about 220~MeV have been reported \cite{Schmidt:2008ev,Karsch:2008fe,Gavai:2005yk,Gavai:2008zr}. }.

A system in thermal equilibrium (for a grand-canonical ensemble) is characterized by its partition function
\begin{eqnarray}
 Z={\rm Tr} \left[ \exp \left(-\frac{H-\sum_i \mu_i Q_i}{T} \right) \right]
\label{eq:partition}
\end{eqnarray} 
where $H$ is the Hamiltonian of the system, and  $Q_i$ and $\mu_i$ denote the conserved charges and the corresponding chemical potentials, respectively. In case of three flavor QCD these are strangeness, baryon-number, and electric charge, or, equivalently, the three quark flavors up, down, and strange. The mean and the (co)-variances are then expressed in terms of derivatives of the partition function with respect to the appropriate chemical potentials\footnote{Although in this article we will mostly concentrate on susceptibilities involving conserved charges, we note that one can define susceptibilities involving any well defined operator. One prominent example is the chiral susceptibility $\chi_{ch}=T/V \, \partial^2/\partial m_q^2 \, \log(Z)$ which characterizes the chiral phase transition in QCD.},
\begin{eqnarray}
 \ave{Q_i} &=& T\frac{\partial}{\partial \mu_i} \log(Z)
\\
\ave{\delta Q_i \delta Q_j} &=&T^2 \frac{\partial^2}{\partial \mu_i \partial \mu_j} \log(Z) \equiv VT \chi_{i,j}
\label{variance}
\end{eqnarray} 
with $\delta Q_i = Q_i - \ave{Q_i}$. Here we have introduced the susceptibilities
\begin{eqnarray}
 \chi_{i,j} = \frac{T}{V} \frac{\partial^2}{\partial \mu_i \partial \mu_j} \log(Z)
\label{eq:susz}
\end{eqnarray} 
which are generally quoted as a measure of the (co)-variances. The diagonal susceptibilities, $\chi_{i,i}$, are a measure for the fluctuations of the system, whereas the off-diagonal susceptibilities, $\chi_{i,j} \, i\neq j$, characterize the correlations between the conserved charges $Q_i$ and $Q_j$. We note that the susceptibilities are directly related to the well known cumulants in statistics \cite{Abramowitz}.

One can define and study higher order susceptibilities or cumulants, by differentiating multiple times with respect to the appropriate chemical potentials
\begin{eqnarray}
\chi^{(n_i,n_j,n_k)} \equiv \frac{1}{V T} \frac{\partial^{n_i} }{\partial (\mu_i/T)^{n_i}} \frac{\partial^{n_j} }{\partial (\mu_j/T)^{n_j}} \frac{\partial^{n_k} }{\partial (\mu_k/T)^{n_k}} \log Z.
\label{eq:higher_susz}
\end{eqnarray}

\begin{figure}[tb]
\begin{center}
\begin{minipage}[c][9.5cm][c]{5.0cm}
\begin{center}
\epsfig{file=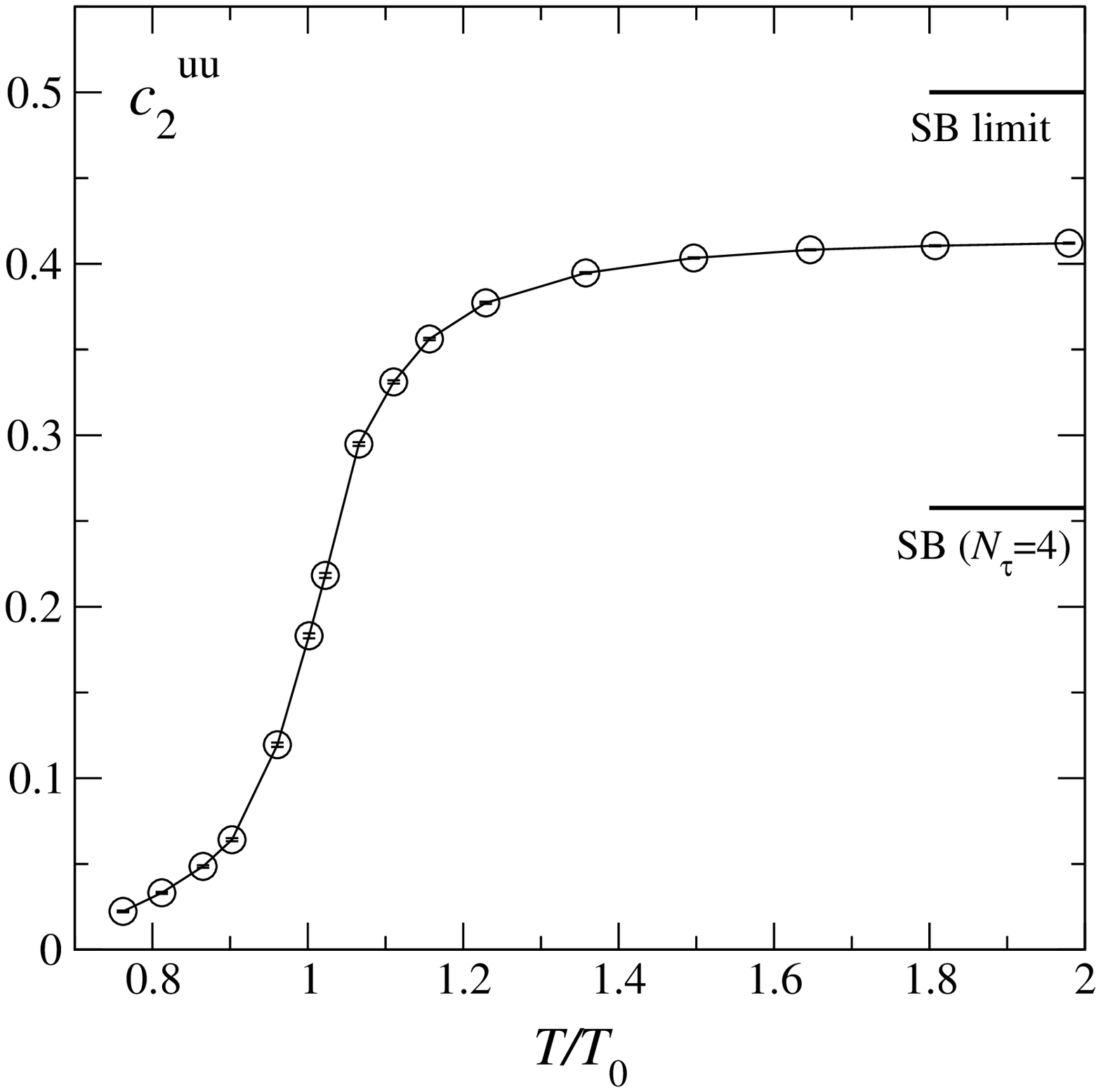, width=5.0cm}\\[-1mm]
\epsfig{file=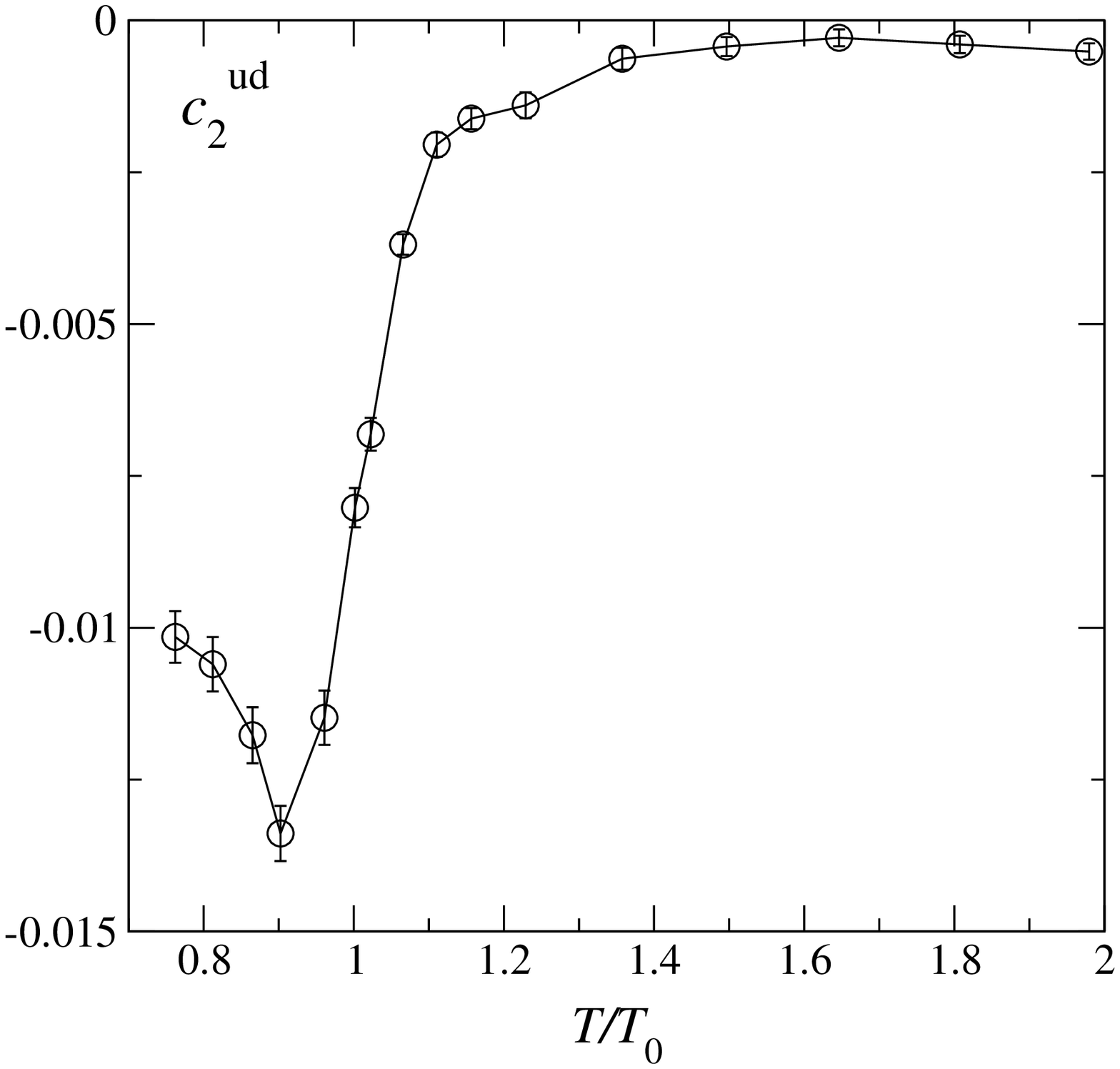, width=5.0cm}\\[-1mm]
\end{center}
\end{minipage}
\begin{minipage}[c][9.5cm][c]{5.0cm}
\begin{center}
\epsfig{file=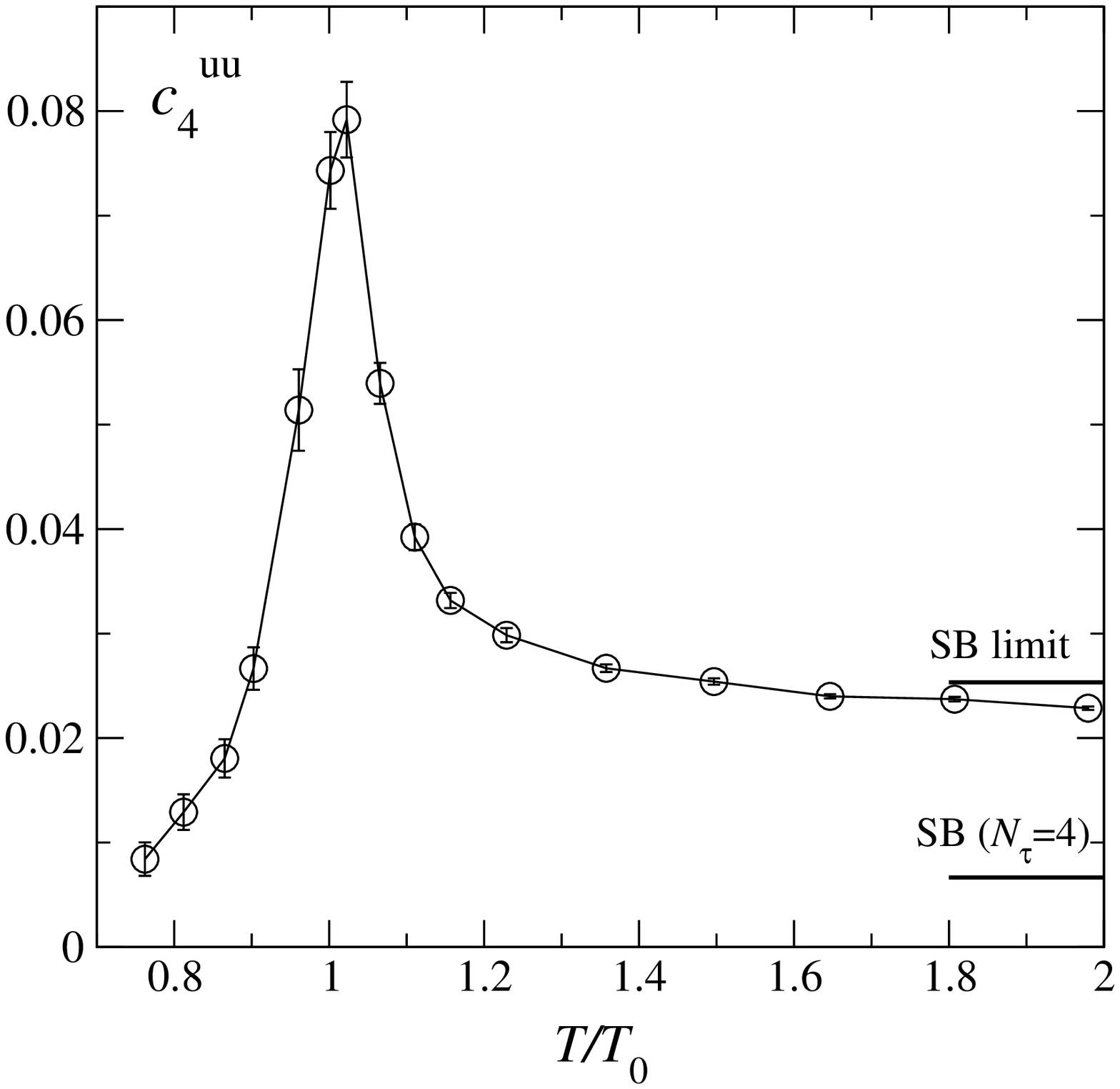, width=5.0cm}\\[-1mm]
\epsfig{file=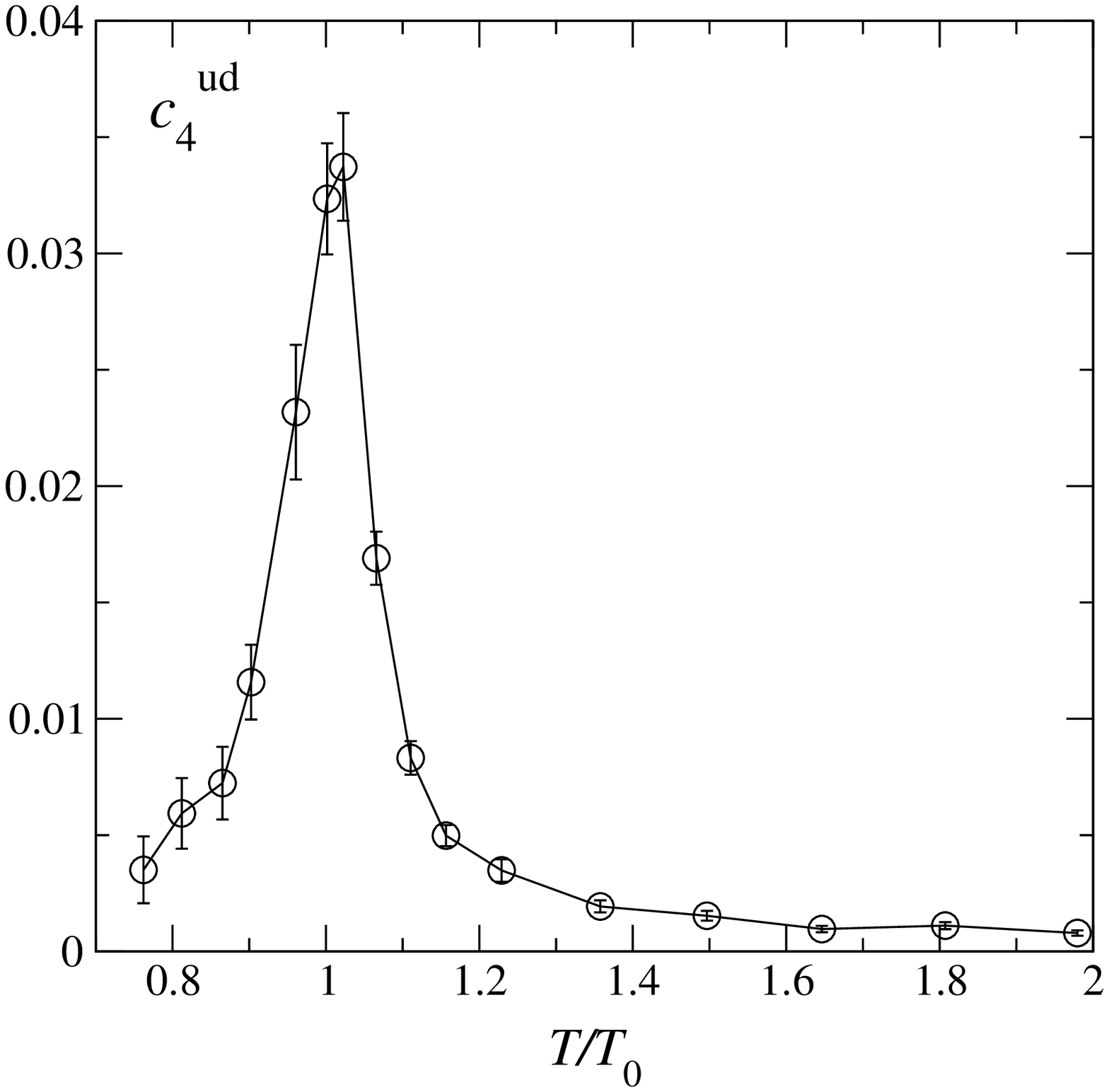, width=5.0cm}\\[-1mm]
\end{center}
\end{minipage}
\begin{minipage}[c][9.5cm][c]{5.0cm}
\begin{center}
\epsfig{file=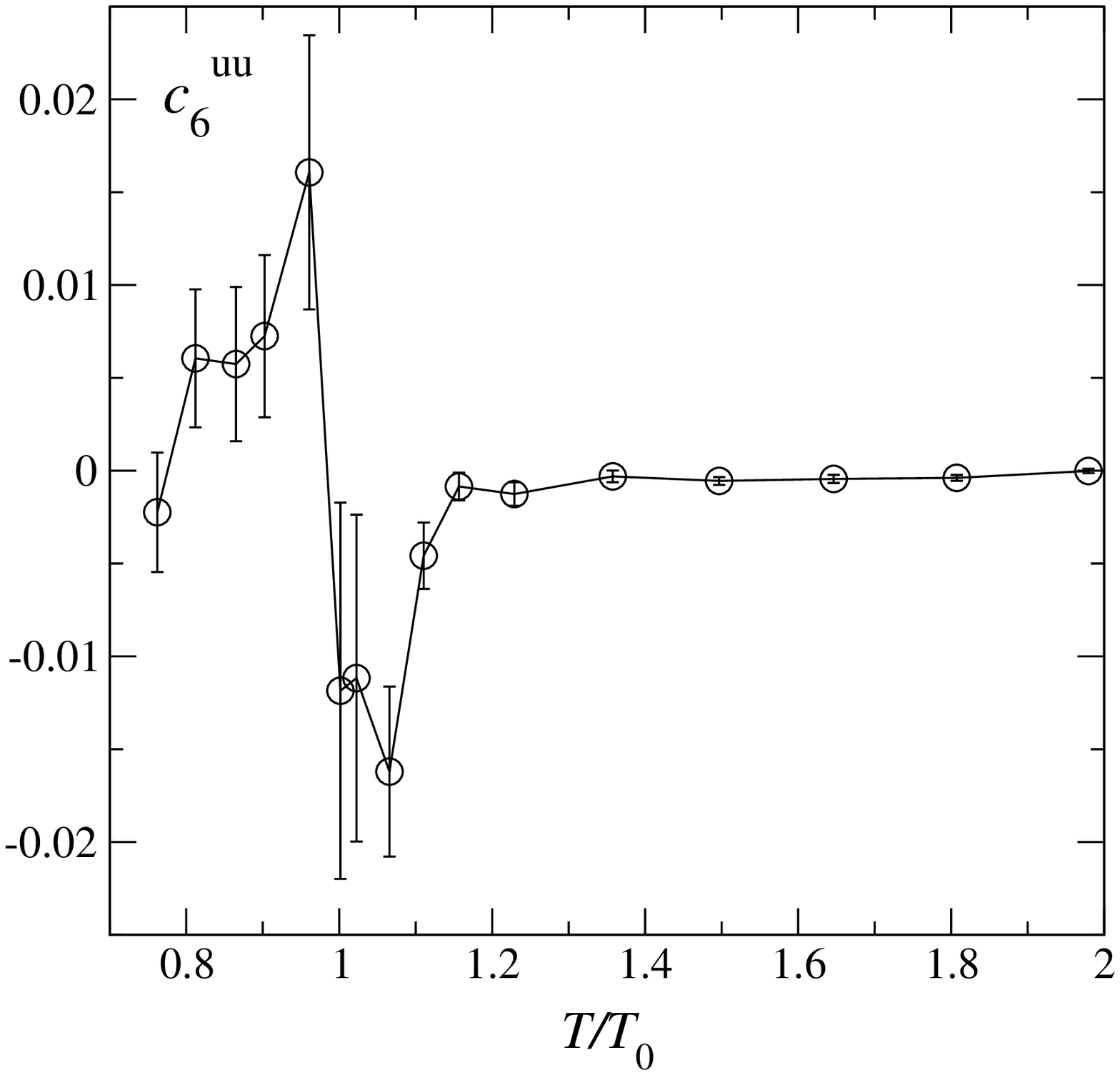, width=5.0cm}\\[-1mm]
\epsfig{file=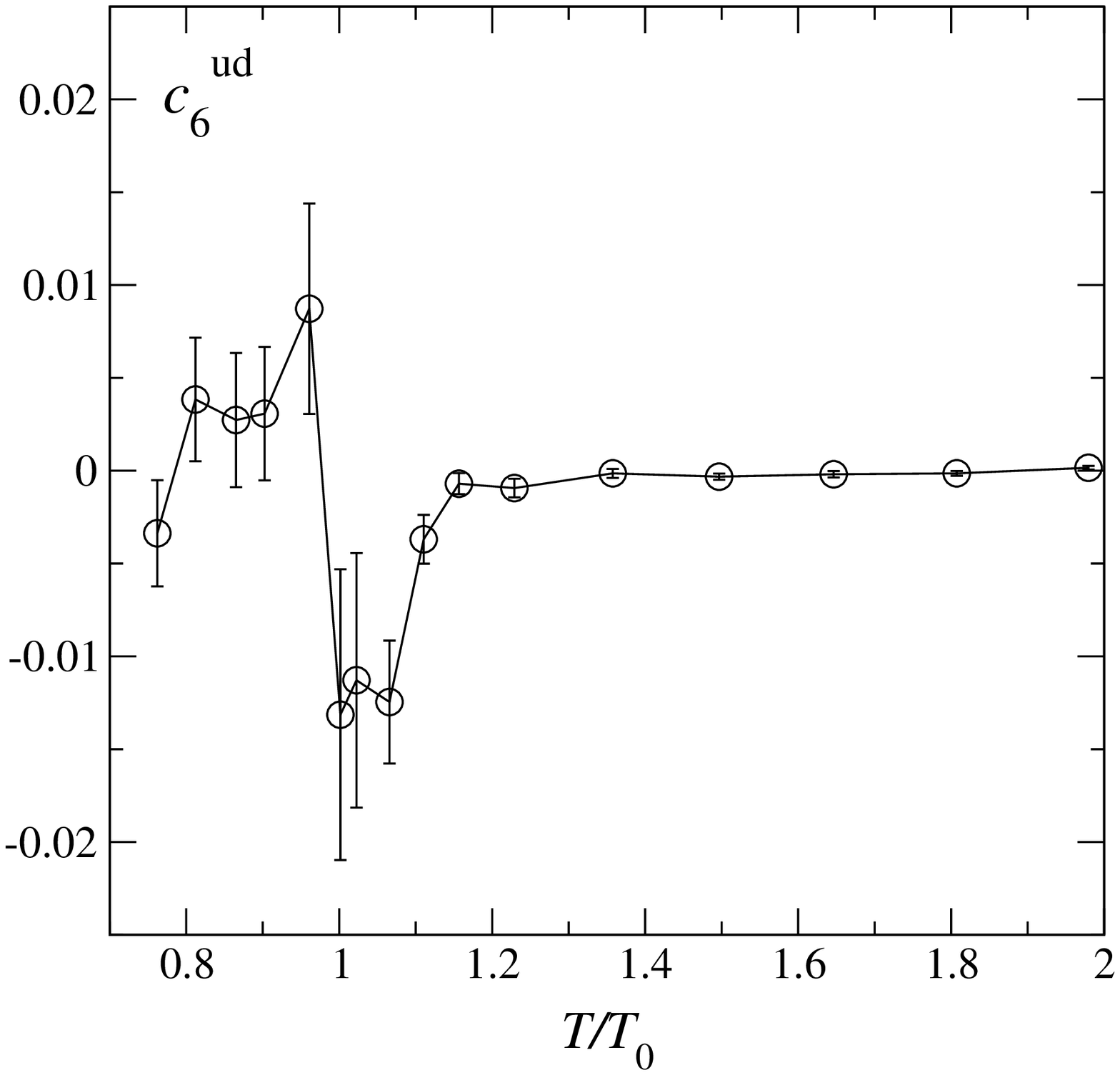, width=5.0cm}\\[-1mm]
\end{center}
\end{minipage}
\caption{The Taylor expansion coefficients $c_n^{uu}$ (upper row) of diagonal and 
$c^{ud}_n$ (lower row)  of non-diagonal susceptibilities from \cite{Allton:2005gk}. Their relation with the susceptibilities is explained in the text, Eq. \ref{eq:taylor_susz}.}
\label{fig:karsch_susz}
\end{center}
\end{figure}

Higher order cumulants up to the sixth \cite{Gavai:2005yk,Allton:2005gk} and even eighth \cite{Gavai:2008zr} order have been calculated in Lattice QCD which, as we will discuss in section \ref{sec:conserved}, provide useful information about the properties of the matter above the critical temperature. In Fig. \ref{fig:karsch_susz} we show the Taylor expansion coefficients obtained in two flavor LQCD \cite{Allton:2005gk} which are proportional to the susceptibilities defined in Eq. \ref{eq:higher_susz}. The upper row in Fig. \ref{fig:karsch_susz} shows the flavor-diagonal susceptibilities up to sixth-order,
\begin{eqnarray}
 c_2^{u,u} &=& \frac{1}{2 T^2}\chi_{u,u} \non
c_4^{u,u}&=&\frac{1}{24 T^2} \frac{\partial^2}{\partial (\mu_q/T)^2} \chi_{u,u}
\non
c_6^{u,u}&=&\frac{1}{144 T^2} \frac{\partial^4}{\partial (\mu_q/T)^4} \chi_{u,u}.
\label{eq:taylor_susz}
\end{eqnarray}
whereas the lower row shows the flavor off-diagonal susceptibility
\begin{eqnarray}
 c_2^{u,d}=\frac{1}{2 T^2}\chi_{u,d}
\end{eqnarray}
and its derivative with the quark number chemical potential $\mu_q=\mu_u+\mu_d$
\begin{eqnarray}
 c_4^{u,d}&=&\frac{1}{24 T^2} \frac{\partial^2}{\partial (\mu_q/T)^2} \chi_{u,d}
\non
c_6^{u,d}&=&\frac{1}{144 T^2} \frac{\partial^4}{\partial (\mu_q/T)^4} \chi_{u,d}.
\end{eqnarray}

We should point out that the results shown in Fig. \ref{fig:karsch_susz} are based on simulation with rather large quark masses. Recently, new results for three flavor QCD with almost physical light quark masses have been reported \cite{Schmidt:2008ev,Schmidt:2008new}. In this case, the second order susceptibilities are consistent with the Stefan-Boltzmann limit of free, uncorrelated massless quarks right above the transition temperature $T_c$, while the results shown here (upper left panel) exhibit a 20\% suppression. The phenomenological consequences of this and other physics interpretations of these susceptibilities will be discussed in the following section.  

As already mentioned, susceptibilities are related to intergrals of equal time correlation functions of the 
appropriate charge-densities. Here we will restrict ourselves to the second order susceptibilities keeping in mind that the higher order susceptibilities can also be expressed in terms of appropriate (higher order) correlation functions.

Consider a density fluctuation $\delta \rho_i(x) = \rho_i(x) - \bar{\rho_i}$ at location $x$, where $\bar{\rho}_i$ denotes the spatially averaged density of the charge $Q_i $. Then the susceptibilities are given by the following integral over the density-density correlation functions:
\begin{eqnarray}
 \chi_{i,j} = \frac{1}{VT}\int d^3x d^3y \ave{\delta \rho_i (x) \, \delta \rho_j (y)} = \frac{1}{T} \bar{\rho_i} \delta_{i,j} + \frac{1}{T} \int d^3 r C_{i,j}(r).
\label{eq:corr_susz}
\end{eqnarray} 
The correlation functions 
\begin{eqnarray}
 C_{i,j}(\vec{r}) =  \ave{\delta \rho_i (\vec{r}) \, \delta \rho_j (0)} - \bar{\rho_i}\delta_{i,j} \delta(\vec{r}) \sim \frac{\exp\left[-r/\xi_{i,j}\right]}{r}
\end{eqnarray} 
are characterized by typical correlation lengths $\xi_{i,j}$. The correlation length provides a measure for the strength and type of the correlation. For example, in case of a second order phase transition the correlation length diverges with a characteristic critical exponent, usually denoted as $\nu$. 

To illustrate this point let us first consider the case of a classical ideal gas. This will also serve as useful reference for comparison with LQCD results.  Since a classical ideal gas has no correlations, by construction its correlation functions  vanish, $C_{\rm ideal \, gas}= 0$, and the susceptibilities are given by the first, local term in Eq. \ref{eq:corr_susz}, $\sim \bar{\rho_i} \delta_{i,j}$, implying that all co-variances vanish. As a consequence, the fluctuations are proportional to the number of particles in the system, and thus grow linearly with the system size, $V$.
\begin{eqnarray}
  \ave{(\delta Q_i)^2} \sim V
\end{eqnarray}

The more relevant case concerning the QCD critical point corresponds to a second order phase transition. In this case, the correlation length diverges at the critical temperature
\begin{eqnarray}
 \xi \sim \absol{T-T_c}^{-\nu}
\end{eqnarray} 
where $\nu>0$ is relevant critical exponents characterizing a second order phase transition in a given universality class \cite{Landau_Stat}. In this case, the volume dependence of the susceptibilities is governed by the integral of the correlation function
\begin{eqnarray}
 \chi_{i,j} \sim \xi^2 \sim V^{2/3}
\end{eqnarray}  
so that the fluctuations grow like\footnote{The correct scaling of the susceptibility with the volume actually involves the critical exponents, $\chi \sim V^{\gamma/(3\nu)}$. Our example here is correct for so called mean field exponents. For details see \cite{Landau_Stat}.}
\begin{eqnarray}
 \ave{(\delta Q_i)^2} \sim V^{5/3}, \,\,\,\,\rm second \, order.
\end{eqnarray} 

In case of a first order transition we have co-existence of phases with different densities, and the correlation function is a constant, $C(r) = const \neq 0$. Consequently, the fluctuations scale like
\begin{eqnarray}
 \ave{(\delta Q_i)^2} \sim V^2,  \,\,\,\,\rm first \, order.
\end{eqnarray} 

Most other systems, including systems with a cross over, such as QCD at vanishing chemical potential \cite{Aoki:2006we}, will exhibit a finite correlation length. Consequently, the susceptibility is independent of the  volume, and the fluctuations scale linearly with the volume, just as in the case of an ideal gas
\begin{eqnarray}
 \ave{(\delta Q_i)^2} \sim V,  \,\,\,\,\rm no\, phase - transition.
\end{eqnarray}

In principle, one could utilize the above volume scaling of the fluctuations in heavy-ion experiments by studying the system size dependence of, e.g., baryon-number fluctuations. However, in case of the second order phase-transition, the phenomenon of critical slowing down limits the actual size of correlation length due to the finite life-time of the system created in these collisions. A maximum correlation length of $\xi \sim 2.5 \, \rm fm$ has been estimated in ref. \cite{Berdnikov:1999ph} which is much smaller than the typical size of a system created in these reactions. Consequently, such a system would just behave like any other with a constant correlation-length and, therefore, would not exhibit the system size dependence discussed above. 

After having introduced the basic concepts and  definitions for fluctuations in a thermal system, in the following section we will discuss, in a few examples, what Lattice QCD results on the various susceptibilities can tell us about the structure of the matter above the transition temperature. These examples will also guide us in developing observables which can and have been measured in heavy-ion experiments.

  \section{Fluctuations and correlations of conserved charges}
\label{sec:conserved}
After having introduced the basic concepts and definitions for fluctuations and correlations in a thermal system, in this section we will discuss some of the physical implications of the results obtained from Lattice QCD. Before we do that, however, let us start with some general considerations. First, let us address the seemingly contradictory notion of fluctuations of \emph{conserved} charges. Obviously, if we look at the entire system, none of the conserved charges will fluctuate. However, by studying a sufficiently small subsystem, the fluctuations of conserved quantities become meaningful. The small system may exchange conserved quanta with the rest of the system. This is similar to the assumptions which govern a thermal system in the grand-canonical ensemble, and Lattice QCD calculations are carried out in this ensemble\footnote{The treatment and correction for total charge conservation is non-trivial and needs to be done carefully \cite{Jeon:2003gk} since the variance in a grand-canonical and canonical ensemble differ even for large systems \cite{Begun:2005ah}.}.

\begin{figure}[ht]
\begin{center}
\epsfig{file=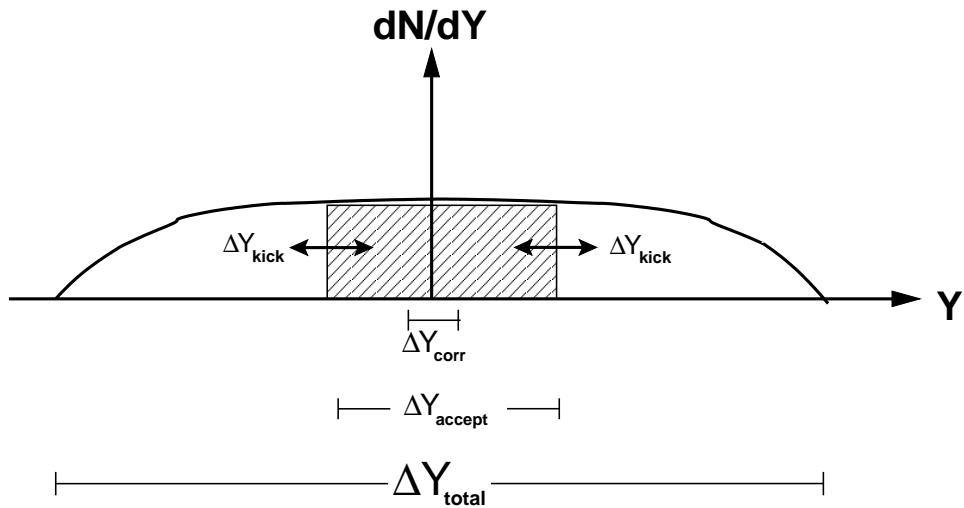,width=0.8\textwidth}
\end{center} 
\caption{The various rapidity scales relevant for charge fluctuations.}
\label{fig:charge_scales}
\end{figure} 

On the other hand, the conservation of the charge ensures that the total charge within the small system can only change by transport of charges through the boundaries leading to rather long equilibration times for charge fluctuations in the hadronic phase \cite{Shuryak:2000pd,Jeon:2003gk,Bleicher:2000ek}. To illustrate this point in the context of  heavy-ion collisions, consider a situation as depicted in Fig.\ref{fig:charge_scales}. The total system corresponds to \emph{all} particles distributed in rapidity $Y$ over a range $Y_{total}$, whereas the small subsystem corresponds to the particles within the accepted rapidity interval $Y_{accept}$. For fluctuations of conserved charges to be a meaningful observable the following scales need to be well separated:
\begin{itemize}
\item The range $\Delta Y_{total}$ for the total charge multiplicity distribution.
\item The interval $\Delta Y_{accept}$ for the accepted charged particles.
\item The charge correlation length $\Delta Y_{corr}$ characteristic to the physics of interest.
\item The typical rapidity shift $\Delta Y_{kick}$ charges receive during and after hadronization.
\end{itemize}
Given these scales, charge fluctuations will be able to tell us about the properties of the early stage of the system, the QGP, if the following criteria are met:
\begin{eqnarray}
\Delta Y_{accept} \gg \Delta Y_{corr}
\label{eq:catch}\\
 \Delta Y_{total} \gg \Delta Y_{accept} \gg \Delta Y_{kick}
\label{eq:keep}
\end{eqnarray} 
The first criterion, Eq. \ref{eq:catch}, is necessary in order to be sensitive to the relevant physics, whereas the second one, Eq. \ref{eq:keep}, ensures that total charge conservation does not suppress the signal and that the signal survives hadronization and the hadronic phase. In particular the condition $\Delta Y_{accept} \gg \Delta Y_{kick}$ is unique to conserved charges. For the charge of the system to change, charges need to be transported through the boundaries of the system. And if the condition $\Delta Y_{accept} \gg \Delta Y_{kick}$ is satisfied it requires many kicks to change the charge of the system. Consequently the relaxation time into a new equilibrium state may be very long, depending on how well the scales are separated. If the charges would not be conserved, on the other hand, they could be produced anywhere within the system leading to a much more rapid equilibration. These considerations have been explored in detail in \cite{Shuryak:2000pd} on the basis of some generalized diffusion equations.

Hence, for sufficiently high bombarding energies and a sufficiently large acceptance, one should be able to see ``back'' into the QGP, provided that hadronization leads to a limited distribution of charge kicks, such that one can assign a typical scale $\Delta Y_{kick}$. The charge kick due to re-scattering in the hadronic phase can be estimated using transport models and one finds \cite{Shuryak:2000pd} $ \Delta Y_{kick} \simeq 2$, close to a naive estimate from $\rho$ decays, which gives $ \Delta Y_{kick} \simeq 1.5$. An estimate of the charge transport during hadronization is very difficult on the other hand. If, for example, the charge kicks are distributed according to a power law, the above considerations are invalid, as the charge transport involves the entire system. This might very well be the reason for the results of \cite{Haussler:2007un,Haussler:2008gx} where in a quark molecular dynamics model, the charge fluctuations quickly adjust to the hadron gas value during hadronization. We note, that in this model string breaking during hadronization is not taken into account, allowing for charge transport over large rapidity intervals due to the linearly increasing potential among the quarks. 

Conversely, at low bombarding energies the necessary separation of scales is not necessarily guaranteed, and the interpretation of fluctuations and correlations of conserved charges becomes rather difficult, as charge conservation effects dominate the observables \cite{Zaranek:2001di}. For example, at top SPS energies, $E_{Beam}=160 \,\rm AGeV$ the width in rapidity of the total multiplicity distribution is $\Delta Y_{total} \simeq 2$ which is comparable to the charge ``kick'' $\Delta Y_{kick} \geq 1.5$ due to the decay of the $\rho$  and Delta resonance. 

After these general considerations about the condition under which the study of charge fluctuations in heavy-ion collisions are meaningful let us discuss some specific examples which have been discussed in the literature. 

The fluctuations of conserved charges in heavy-ion collisions was first discused in the context of (electric) net-charge fluctuations \cite{Jeon:2000wg,Asakawa:2000wh}. Here the simple observation was that charge fluctuations per entropy should scale with the \emph{square} of the electric charge of the charge carrying particles and, consequently, they should be sensitive to the fractional charges of the quarks.
For simplicity, consider a classical ideal gas of positively and negatively charged 
particles of charge $\pm q$. The fluctuations
of the total charge contained in a subsystem of $N$ particles is then given by
\begin{eqnarray}
\ave{\delta Q^2} 
&=& q^2 \ave{(\delta N_+ - \delta N_-)^2}
\non 
&=& q^2 \left[ \ave{\delta N_+^2} + 
\ave{\delta N_-^2}
\right]
\non
&=& q^2\left[\ave{N_+}+\ave{N_-}\right]= q^2 \ave{N_{ch}}. 
\end{eqnarray}
since $ \ave{\delta N_- \delta N_+}=0$ and $\ave{\delta N^2} = \ave{N}$ for an ideal gas\footnote{Corrections due to quantum statistics are small, of the order of one percent for the temperatures under consideration \cite{Bertsch:1994qc,Jeon:2003gk}.}. For a two flavor  \QGP this translates into
\be
\ave{\delta Q^2_{q}} = Q_u^2 \ave{N_u} + Q_d^2 \ave{N_d}  
= \frac{5}{9} \ave{N_q}
\label{eq:ch1:dq_qgp}
\ee
where $N_q = N_u = N_d$ denotes the number of quarks {\em and} anti-quarks.
For a pion gas, on the other hand, we have
\be
\ave{\delta Q_\pi^2} = \ave{N_{\pi^+}} + \ave{N_{\pi^-}}.
\label{eq:ch1:dq_pi_gas}
\ee
Dividing both by the entropy of massless classical particles 
\be
S = 4 \ave{N}
\label{eq:ch1:classical_entropy}
\ee
weighted by the appropriate number of degrees of freedom, 
three in case of the pion gas and 40 in case of the QGP, we get
\begin{eqnarray}
\frac{\ave{\delta Q_q^2}}{S_{\rm QGP}} = \frac{1}{24} 
\end{eqnarray}
for a 2-flavor  \QGP  whereas for a pion gas we obtain
\be
\frac{\ave{\delta Q_\pi^2}}{S_\pi} = \frac{1}{6}.
\ee
Thus, the charge fluctuations per degree of freedom 
in a \QGP are a factor of four 
smaller than those in a pion gas. Hadronic resonances, which
constitute a considerable fraction of a hadron gas, reduce  the result for
the pion gas by about 30 \% \cite{Jeon:2000wg,Jeon:1999gr}, leaving still a
factor 3 signal for the existence of the Quark Gluon Plasma. We note that similar arguments can be made for baryon number fluctuations \cite{Asakawa:2000wh}, where one is sensitive to the fractional baryon number of the quarks. However, in this case an actual measurement would require the detection of neutrons, which is rather difficult.

As already discussed in the previous section both the charge fluctuations and entropy can be calculated using Lattice QCD. For the two flavor calculations with rather large quark masses \cite{Allton:2005gk,Karsch:2003jg}, above the transition temperature both the charge fluctuation and the entropy deviate by about 20\% from that of a non-interacting gas of quarks and gluons, so that the ratio agrees with our estimates. In case of the more recent three-flavor calculation, the charge fluctuations are very close to the free gas limit \cite{Schmidt:2008ev} and the entropy deviates by at most 10\% \cite{Karsch:2008fe}, so that the ratio again is in good agreement with our simple estimate. 

Unfortunately, at present lattice results are not
available for this ratio much below the critical temperature. Here one has to
resort to hadronic model calculations. This has been done in
\cite{Doring:2002qa,Eletsky:1993hv} using either a virial expansion, a chiral
low energy expansion or an explicit diagrammatic calculation. In all cases, interactions slightly increase the
ratio as compared to a system of non-interacting hadrons, thus enhancing the signal for
the Quark Gluon Plasma.

Finally we should note, that present measurements at RHIC and SPS have not found the small net-charge fluctuations predicted for a QGP \cite{Jeon:2000wg,Asakawa:2000wh} and are rather consistent with a resonance gas. 

While the charge fluctuations involve the diagonal susceptibilities, let us next discuss an example where the off-diagonal susceptibilities play an essential role. In ref. \cite{Koch:2005vg} the following simple observation was made: If above the transition temperature the quarks can be well described by uncorrelated quasiparticles, then any object which carries strangeness necessarily is a quark, and thus carries baryon number. Contrast this with a hadron gas, where kaons are mesons which carry strangeness. Thus one expects  baryon number and strangeness to be stronger correlated in a \QGP than in a hadron gas. To quantify this, let us introduce the following ratio \cite{Koch:2005vg} 
\begin{equation}
C_{BS} \equiv -3\frac{\sigma_{BS}}{\sigma_S^2} =
-3 \frac{\ave{BS}-\ave{B}\ave{S}}{\ave{S^2} - \ave{S}^2}
= -3 \frac{\ave{BS}}{\ave{S^2}} .
\label{eq:cbs}
\end{equation}
In terms of quark flavors the correlation coefficient $C_{BS}$ can be written as
\begin{eqnarray}
C_{BS}= -3 \frac{\ave{BS}}{\ave{S^2}} = \frac{\ave{(u+d+s)(s)}}{\ave{s^2}}=1+\frac{\ave{us}+\ave{ds}}{\ave{s^2}}=1+\frac{\chi_{us}+\chi_{ds}}{\chi_{ss}}
\label{eq:cbs_flavor}
\end{eqnarray}
since the baryon number of a quark is $1/3$ and the strangeness of a strange quark is minus one. 
Note the the quark operators $u,d,s$ here represent the \emph{net}-quarknumber of a given flavor, i.e. $\ave{u}\equiv \ave{u-\bar{u}}$ etc. Also, to obtain the last expression in Eq. \ref{eq:cbs_flavor}, we used the definition for the susceptibilities, Eq. \ref{eq:susz} from the previous section.
For a simple quark-gluon plasma, or more generally, for uncorrelated quark flavors, we have
\begin{eqnarray}
 \ave{us}=\ave{ds}=0
\end{eqnarray}
and hence 
\begin{eqnarray}
 C_{BS}=1.
\end{eqnarray}
In contrast, a gas of uncorrelated hadron resonances gives  
\begin{equation}
C_{BS}\ \approx\ 3 \frac{\ave{\Lambda} + \ave{\bar{\Lambda}} + \dots\,
+3 \ave{\Omega^-} + 3\ave{\bar{\Omega}^+}}{ 
\ave{K^0} + \ave{\bar{K}^0} + \dots\,
+9 \ave{\Omega^-} + 9\ave{\bar{\Omega}^+}}.
\label{eq:simple}
\end{equation} 
Here the numerator receives contributions
only from (strange) baryons,
while the denominator receives contributions also from (strange) mesons.
As a result, $C_{BS}=0.66$ for $T=170~{\rm MeV}$ 
and $\mu_B=0$.

\begin{figure}
\begin{center}
\includegraphics[width=0.45\textwidth]{cbs_mu.eps}
\includegraphics[width=0.45\textwidth]{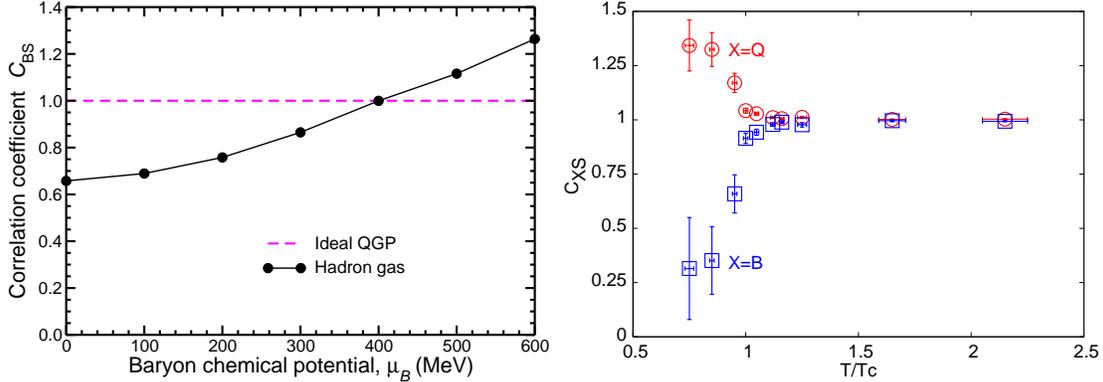}
\end{center}
\caption{Left panel: The correlation coefficient $C_{BS}$
for a hadron gas along the empirical chemical freeze out line as determined in \cite{Braun-Munzinger:2003zd} (full line).  The dashed line corresponds to and ideal quark-gluon plasma. The figure is from \cite{Koch:2005vg}. Right panel:
$C_{BS}$ (squares) and $C_{QS}$ (circles) as a function of temperature from Lattice QCD \cite{Gavai:2005yk}. }
\label{fig:cbs}
\end{figure}

On the other hand, at very high $\mu_B$ and low $T$, where strangeness is
carried exclusively by Lambdas and Kaons, $C_{BS}\approx\frac{3}{2}$, since $\ave{\Lambda}=\ave{K}$ due to strangeness neutrality.
This significant dependence of $C_{BS}$ on the hadronic environment
is in sharp contrast to the simple quark-gluon plasma
where the correlation coefficient remains strictly one
at all temperatures and chemical potentials. In the left panel of  Fig. \ref{fig:cbs} we show the result for $C_{BS}$ for both an ideal quark-gluon plasma as well as a hadron gas along the empirical chemical freeze-out line, as determined e.g. in  \cite{Braun-Munzinger:2003zd}.

The correlation coefficient $C_{BS}$ has been evaluated in Lattice QCD \cite{Gavai:2005yk}. The result is shown in the right panel of Fig. \ref{fig:cbs} (squares). We see that right above the transition temperature $C_{BS}=1$ indicating a system of uncorrelated (quasi)-quarks. Also shown in this Figure is the correlation between strangeness and the electric charges
\begin{eqnarray}
 C_{QS}=3 \frac{\ave{QS}}{\ave{S^2}}=\frac{\ave{(2u - d -s)(-s)}}{\ave{s^2}}=1-\frac{2\chi_{us}-\chi_{ds}}{\chi_{ss}}
\end{eqnarray}
which also approaches a value of $C_{QS}=1$ right above the transition temperature $T_c$, again indicating uncorrelated quark flavors\footnote{A very recent lattice calculation \cite{Schmidt:2008new} using almost physical quark masses shows a somewhat softer transition, reaching the limit of uncorrelated quarks at $T/T_c \simeq 1.3$}. Both results, $C_{BS}\simeq 1$ and $C_{QS}\simeq 1$ above $T_c$ are equivalent, as both result from the fact that the flavor off-diagonal susceptibilities are much smaller than the flavor diagonal ones
\begin{eqnarray}
 \chi_{us}\ll\chi_{ss}\\
\chi_{ds}\ll\chi_{ss}.
\end{eqnarray}
The correlation coefficient $C_{QS}$ may be the better suitable for experimental investigation, since, contrary to $C_{BS}$, it does not involve the measurement of neutrons. There may be even better combinations of quantum numbers as discussed in \cite{Majumder:2006nq}. However they all rely on the fact that the ratio of flavor-off diagonal to flavor diagonal susceptibilities is small above the transition temperature.

\begin{figure}
\centerline{\includegraphics[height=7cm,angle=-90]{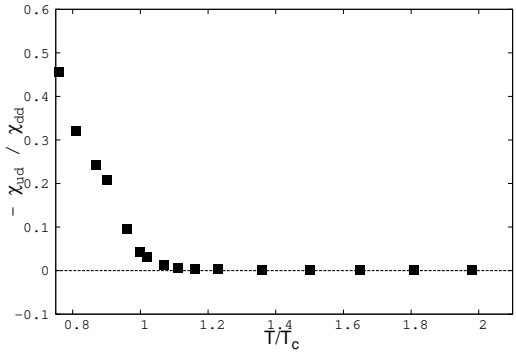}}
\caption{Negative ratio of off-diagonal to diagonal susceptibility as a function of temperature in 2-flavor unquenched Lattice QCD. Lattice results are from \cite{Allton:2005gk}. Figure adapted from \cite{Koch:2006ek}. }
\label{fig:x_ud}
\end{figure}

A similar argument can be made using only light flavors. In this case one would look at something like baryon-number -- ``up-ness'' correlations. Again the relevant quantity would be the ratio of flavor off-diagonal to flavor diagonal susceptibilities, $\chi_{ud}/\chi_{dd}$ and uncorrelated flavors would correspond to
\begin{eqnarray}
 \frac{\chi_{ud}}{\chi_{dd}} \ll 1.
\end{eqnarray}
In Fig.\ref{fig:x_ud} we show this ratio for the light quarks based on the Lattice data of ref. \cite{Allton:2005gk}. The diagonal and off-diagonal susceptibilities are also shown separately in Fig.\ref{fig:karsch_susz}.
We find that the ratio drops rapidly as we approach the transition temperature $T/T_c = 1$ and for temperatures $T/T_c \gtrsim 1.2$, the ratio is consistent with zero, indicating independent quark flavors. This behavior is consistent with a rapid melting of the hadronic states in the transition region. It also suggests that the system above $T_c$ should be well described in terms of independent quasi particles. And indeed, many of the lattice results, including the susceptibilities can be reproduced in quasiparticle \cite{Peshier:2002ww} and mean field models \cite{Ratti:2007jf}.

\begin{figure}
\begin{center}
 \includegraphics[width=0.45\textwidth]{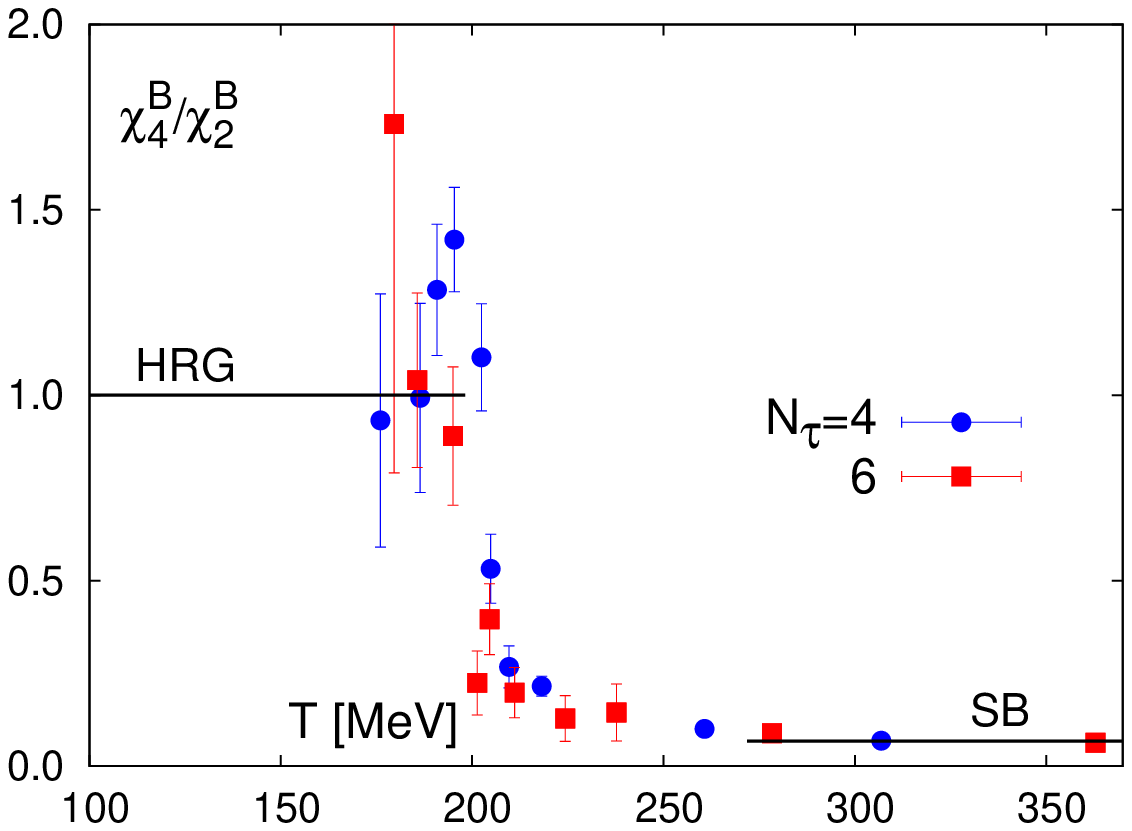}
 \includegraphics[width=0.45\textwidth]{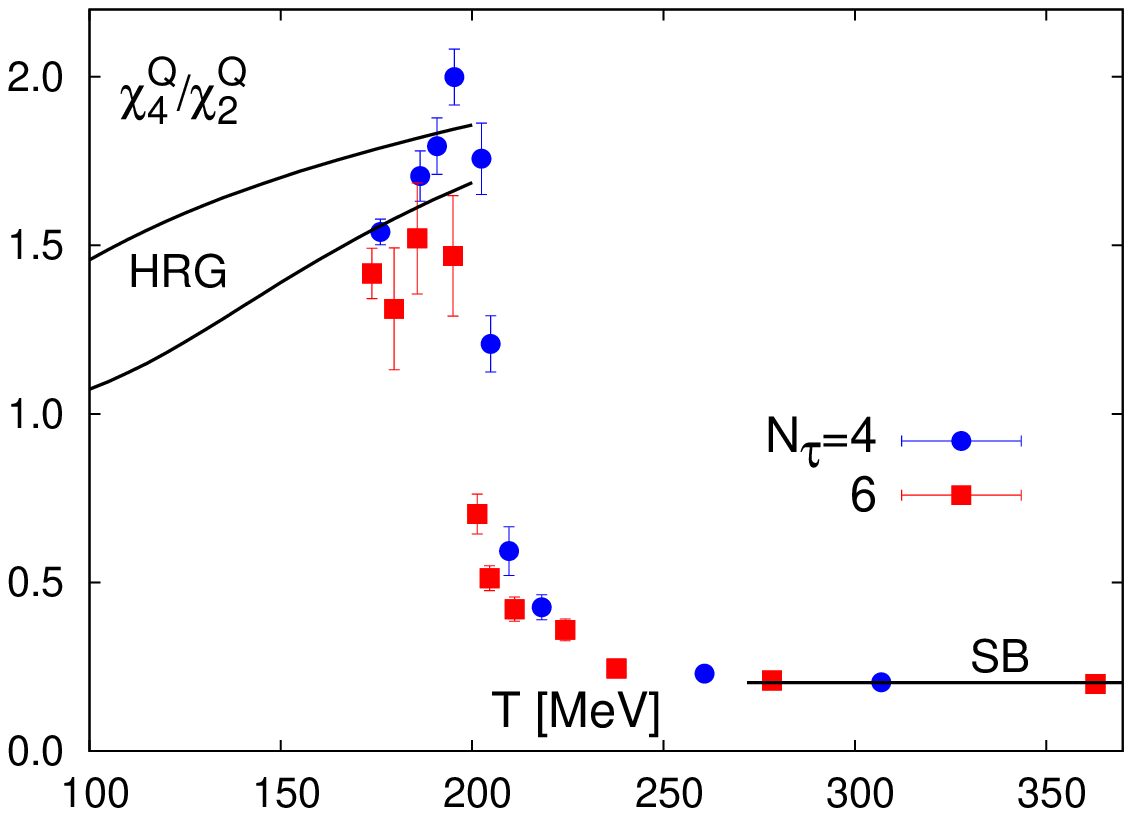}
\end{center}
\caption{Ratio of fourth order to second susceptibilities for baryon number (left panel) and electric charge (right panel). The lines at low temperature, labeled ``HRG'' indicate the results for a hadron gas. The limit of free quarks is shown on the right of each plot and denoted by ``SB''. The two hadron gas lines on the left panel represent a hadron gas calculation with (upper) and without pions (lower). Figure adapted from \cite{Schmidt:2008new}. }
\label{fig:schmidt_b4}
\end{figure}

Further evidence for independent quasi-quarks is found in the higher susceptibilities. In \cite{Ejiri:2005wq} the authors looked at the ratio of the fourth order susceptibilities (cumulants) over the second order susceptibilities for baryon number and net charge
\begin{eqnarray}
R^{(B,Q)}_{4,2} \equiv \frac{\chi^{(4)}_{B,Q}}{\chi^{(2)}_{B,Q}}
\label{eq:susz42}
\end{eqnarray}
where we used the notation of Eq.\ref{eq:higher_susz}. In Fig.\ref{fig:schmidt_b4} we show lattice results from a recent calculation with three flavors and almost physical light quark masses \cite{Schmidt:2008ev,Schmidt:2008new}. On the left panel we have the ratio $R^B_{4,2}$ for baryons and on the right panel the ratio $R^Q_{4,2}$ for the electric charge. In both figures the full lines indicate estimates for a hadron gas (labeled ``HRG'') at low temperatures and  the limit of non-interacting quarks at high temperatures, (labeled ``SB). In addition, the left panel shows two estimates for the hadron gas, one with (higher) and one without pions (lower). We note that for both ratios the peak close to the transition temperature softens considerably when going to smaller lattice spacings, i.e, from $N_\tau=4$ to $N_\tau=6$.  

Let us now concentrate on the left panel, where the ratio for the baryon number susceptibilities is shown. This result can be easily understood in terms of hadrons on the low temperature side and independent quarks on the high temperature side. Consider a classical ideal gas of particles with baryon number $b$. Then we have
\begin{eqnarray}
\chi^{(2)}_{B} &=& b^2 \lb \chi^{(2)}_N + \chi^{(2)}_{\bar{N}} \rb\\
\chi^{(4)}_B &=& b^4 \lb \chi^{(4)}_N +\chi^{(4)}_{\bar{N}} \rb
\end{eqnarray}
 where $\chi^{(2)}_N$ and $\chi^{(2)}_{\bar{N}}$ are the particle-number cumulants for particle and anti-particles, respectively.
\begin{eqnarray}
 \chi^{(2)}_N &=& \ave{N^2}-\ave{N}^2=\ave{N}\\
 \chi^{(4)}_N &=& \ave{N^4}-3\ave{N^2}=\ave{N}
\label{eq:chi_4}
\end{eqnarray}
Consequently
\begin{eqnarray}
 R^B_{4,2}=b^2
\end{eqnarray}
and since all baryons in the hadronic phase have baryon number $|B_{hadronic}|=1$ and all quarks have baryon number $|B_{quark}|=1/3$ we get for the the ratio of the cumulants the simple results
\begin{eqnarray}
 R^B_{4,2}&=&1; \,\,\,{\rm hadron \,\, phase}\\
 R^B_{4,2}&=&\frac{1}{9}; \,\,\,{\rm QGP}\\
\end{eqnarray}
Since baryons are fermions, one would have to correct the above result for quantum statistics. In case of massless particles this can be done analytically and one would have to multiply the results by a factor of $6/\pi^2 \simeq 0.6$. The effect of quantum statistics for the massive baryons in the hadronic phase can only be evaluated numerically and for baryons with mass $M=1\,\rm GeV$ and a temperature of $T=170 \, \rm MeV$ the correction is less than 1\%.
The results of this simple estimate are also shown in Fig.\ref{fig:schmidt_b4} as the full lines at low (hadron gas) and high (QGP) temperatures.

Again, we find that above $\sim 1.5 \, T_c$ the system is well described by a gas of independent quarks. Thus, the study of correlations and fluctuations reveals insights into the structure of strongly interacting matter above the transition temperature which the equation of state alone could not tell us. In addition, if the latest $N_\tau=6$ with light quarks are indeed what the continuum limit will look like, then we see a rapid change from hadrons to independent quarks, both in the flavor off-diagonal susceptibilities as well as in the fourth order baryon number cumulants. This would lend theoretical support of the rather surprising finding of the so-called quark number scaling at RHIC \cite{Voloshin:2002wa,Abelev:2008ed,Afanasiev:2007tv}, which can be simply understood within a recombination/coalescence picture. In this picture hadrons are formed at $T_c$ by simple phasespace coalescence, which is consistent with the the rather rapid buildup of correlations around $T_c$ seen in the lattice results for the susceptibilities as discussed here. If, on the other hand, the peak in the fourth order baryon-number cumulant just below $T_c$ of the $N_\tau=4$ results is what the continuum limit will be, then yet to be understood strong correlations are at work close to the transition temperature. This in turn could be a sign of the existence of a critical point at finite baryon-number chemical potential, where all the baryon-number cumulants are expected to diverge. The physics of the critical point will discussed in more detail in section \ref{sec:critical}. 

To summarize, in this section we have shown that the study of fluctuations and correlations reveals important insights into the underlying structure of the strongly interacting matter, especially above the transition temperature. How this can be translated into experimentally accessible observables will be the subject of the following section.

 \section{Observables}
\label{sec:observable}
In the previous sections we have concentrated on fluctuations and correlations of thermal systems and have presented a few examples on how to extract interesting physics from them. Now let us turn to actual heavy-ion collisions and let us discuss how to construct meaningful observables involving the various susceptibilities discussed previously. 

One essential difference between a system studied in lattice QCD and a heavy-ion collision is that in the latter case the observables are restricted to correlations in the momenta and quantum numbers of the observed particles. 
Spatial correlations, such as the correlations length,  are accessible only
indirectly via Fourier transforms of momentum space
correlations, and thus limited. The susceptibilities, on the other hand, can be measured since they can be expressed as integrals over either spatial or momentum space correlation functions. Thus, as long as one deals with susceptibilities, i.e (co)-variances, there is a one to one mapping from Lattice QCD results to heavy-ion collisions (assuming that the latter produce a system in thermal equilibrium). The susceptibilities can be extracted from data either by studying event-by-even fluctuations of a given quantity or by measuring and integrating the appropriate multi-particle densities \cite{Bialas:1999tv}.

An additional complication in the experimental study of fluctuations in heavy-ion collisions are unavoidable impact parameter fluctuations. These additional sources for fluctuations may very well mask the fluctuations of interest. In the language of thermal systems, these impact parameter fluctuations correspond to volume fluctuations. Consequently, one should study so called {\em intensive} variables, i.e. variables which do not scale with the volume, such as temperature, energy density, etc. 

To simplify the discussion, let us concentrate on two-particle correlations, which fully characterize fluctuations of Gaussian nature\footnote{In the case of fluctuations of positive definite quantities, such as transverse momentum or energy, the appropriate distribution is a so called Gamma-
distribution \cite{Tannenbaum:2001gs}}. 
Let us begin by defining the number of particles with quantum numbers
$\alpha=\{\alpha_1,\ldots,\alpha_n\}$  in the momentum space interval $(p,p+dp)$ in a given {\em event}\footnote{In this section, we will use the term `event' and `member of the given
ensemble' interchangeably. We also use `event average' and `ensemble average'
interchangeably.}
\be
n_p^\alpha = \frac{dN_{\rm event}^\alpha}{dp} \, dp
\ee 
and its fluctuations 
\be
\delta n_p^\alpha = n_p^\alpha - \ave{n_p^\alpha}
\label{eq:delta_n_p}
\ee
where $\ave{n_p^\alpha}$ denotes the event-averaged multiplicity in the momentum bin $(p,p+dp)$
\begin{eqnarray}
 \ave{n_p^\alpha} = \frac{1}{N_{\rm events}}\sum_{{\rm event}\, i=1}^{N_{\rm events}} \frac{d N^\alpha_i}{dp} \, dp.
\end{eqnarray}
 
In the case of an ideal gas, $\ave{n_p^\alpha}$ takes the form of the Bose-Einstein  or the Fermi-Dirac distribution depending on the spin of the particles.

The mean value of an observable\footnote{Here we use sums over momentum states, as appropriate for a finite box. The conversion to continuum state is a usual, $\sum_p \rightarrow V \int \frac{d^3 p}{(2 \pi)^3}$, where $V$ is the volume of the system.}
\begin{eqnarray}
X &=& \sum_{p,\alpha} x_p^\alpha n_p^\alpha
\label{eq:ch4:X}
\end{eqnarray}
is obtained by averaging over all the events in the ensemble 
\begin{eqnarray}
\ave{X} &=& \sum_{p,\alpha} x_p^\alpha \ave{n_p^\alpha}.
\end{eqnarray}
Note that $X$ defined in this way is an extensive observable and thus subject to volume fluctuations.  

For the discussion of fluctuations and correlations we need to consider two-particle densities. Let us, therefore, define the basic correlator $\Delta_{p,q}^{\alpha,\beta}$
\begin{eqnarray}
\Delta_{p,q}^{\alpha,\beta} \equiv 
\ave{\delta n_p^\alpha \delta n_q^\beta} 
\label{eq:ch4:basic_corr}
\end{eqnarray}
where $\delta n_p^\alpha$ is defined in Eq.(\ref{eq:delta_n_p}). In case of an ideal gas, particles with different quantum numbers and in different momentum states are uncorrelated. Therefore,
\begin{eqnarray}
\Delta_{p,q}^{\alpha,\beta} 
= \delta_{pq} \lb  \prod_i^n \delta_{\alpha_i\beta_i} \rb \, \omega_p^\alpha \,
\ave{n_p^\alpha}
\label{eq:ch4:delta_ideal}
\end{eqnarray}
with
\begin{eqnarray}
\omega_p^\alpha = 1
\end{eqnarray}
for a classical ideal gas, and 
\begin{eqnarray}
\omega_p^\alpha = (1 \pm \ave{n_p^\alpha})
\label{eq:ch4:omega_ideal} 
\end{eqnarray}
for a boson ($+$) and a fermion ($-$) gas (see e.g \cite{reif_book}). In the presence of dynamical correlations the basic correlator $\Delta_{p,q}^{\alpha,\beta}$, Eq.\ref{eq:ch4:basic_corr},  may contain off-diagonal elements in momentum space and/or in the space of the quantum numbers. For example, the presence of a resonance, such as a $\rho_0$, not only correlates the number of $\pi^+$ and $\pi^-$ in the final state due to resonance decay, but also their momenta. In addition, the occupation number $n_p$ in a given momentum interval may be changed, for example, due to collective flow effects.

Any two-particle observable can be expressed in terms of the basic correlator $\Delta_{p,q}^{\alpha,\beta}$ \cite{Bialas:1999tv,Stephanov:1999zu}. For the generic observable $X$ as given by Eq. \ref{eq:ch4:X} the variance is given by
\begin{eqnarray}
\ave{\delta X^2} &=& \sum_{p,q,\alpha,\beta} \Delta_{p,q}^{\alpha,\beta} 
x_p^\alpha x_q^\beta.
\label{eq:basic_variance}
\end{eqnarray}
To illustrate the formalism let us consider the net charge
\begin{eqnarray}
Q  = \sum_{p,\alpha} Q(\alpha)\, n_p^\alpha. 
\label{eq:ch4:Q}
\end{eqnarray}
Here, $Q(\alpha)$ is the charge of the particle with the quantum numbers $\alpha=\{\alpha_1,\ldots,\alpha_n\}$. 
Its variance, i.e. the net charge fluctuations, are given by  (see Eq.\ref{eq:basic_variance})
\begin{eqnarray}
\ave{ \delta Q^2} &=& \sum_{p,\alpha;\,q,\beta} Q(\alpha)\,Q(\beta)\, \Delta_{p,q}^{\alpha,\beta}. 
\label{eq:ch4:delta_Q}
\end{eqnarray}
In case the system contains only particles with (positive or negative) unit charge, the above expression simplifies to\footnote{Note the difference in notation: whereas $\delta n_p$ refers to the fluctuations in the momentum interval $(p,p+dp)$, $\delta N$ refers 
to the fluctuations of the total (integrated) number of particles.}
\begin{eqnarray}
\ave{ \delta Q^2} &=&\sum_{p,q} \lb \Delta^{+,+}_{p,q} + \Delta^{-,-}_{p,q}  -2 \Delta^{+,-}_{p,q} \rb \non
&=&\ave{\delta N_+^2} +\ave{\delta N_-^2} -2 \ave{\delta N_+ \delta N_-}. 
\end{eqnarray}
Note, in the last equation the basic correlator $\Delta_{p,q}^{+,+}$ still contains a sum over all
other quantum numbers, which we have suppressed. 
\begin{eqnarray}
\Delta_{p,q}^{+,+} = \sum_{\alpha_i \ne {\rm charge}, \beta_i \ne {\rm
    charge}}\Delta_{p,q}^{\{+,\alpha_i\},\{+,\beta_i\}}
\end{eqnarray}
and similar for the other combinations, meaning that all charged
particles are included independent of their flavor, spin, etc.

For a thermal system, the basic correlator $\Delta_{p,q}^{\alpha,\beta}$ is directly related to the second order susceptibilities
\begin{eqnarray}
\ave{\delta Q_i \delta Q_j} = \sum_{p,\alpha;\, q,\beta} Q_i(\alpha) Q_j(\beta) \, \Delta_{p,q}^{\alpha,\beta}= V T \chi_{i,j}.
\label{eq:chi_delta_relation}
\end{eqnarray}
Here, $Q_i(\alpha)$ denotes the conserved charge $Q_i$, such as electric charge, baryon number, or strangeness, associated with the quantum number $\alpha=\{\alpha_1,\ldots,\alpha_n\}$. For instance, the baryon-number strangeness correlations discussed in section \ref{sec:conserved} are given in terms of the basic correlator as
\begin{eqnarray}
 \ave{BS}=\sum_{p,\alpha;\, q,\beta} B(\alpha) S(\beta)\, \Delta_{p,q}^{\alpha,\beta}.
\end{eqnarray}
Obviously, all the information revealed by (Gaussian) fluctuations 
is encoded in the basic correlator $\Delta_{p,q}^{\alpha,\beta}$ and the 
fluctuations of different observables expose different ``moments'' and
elements of the  basic correlator. As we will discuss below, all the event-by-event fluctuation observables studied so far, such as ratio fluctuations, mean transverse momentum fluctuations, etc, can be expressed as moments of the basic correlators.   

If the system has no genuine two-particle correlations, then the basic
correlator will be
\be
\Delta_{p,q}^{\alpha,\beta} = \ave{n_p} \delta_{p,q} \delta_{\alpha,\beta}
\label{eq:ch4:no_corr}
\ee
similar to the classical ideal gas, except that $\ave{n_p}$ does not have to follow a Boltzmann
distribution. Using the inclusive single particle spectrum as momentum distribution, the relation Eq.\ref{eq:ch4:no_corr} may be utilized as an estimator of the fluctuations due to finite number (Poisson) statistics \cite{Gazdzicki:1992ri}. A more detailed discussion can be found in \cite{Jeon:2003gk}.

Any correlations, on the other hand will lead to a non Poisson behavior of the basic correlator. The ideal quantum gases discussed above are examples where the symmetry of the wavefunction introduces 2-particle correlations which either reduce (Fermions) or enhance (Bosons) the fluctuations. Also global conservation laws such as charge, energy, etc. will affect the fluctuations. 

As mentioned in the beginning of this section, in order to avoid contributions
from volume / impact parameter fluctuations it is desirable to study so
called intensive quantities, i.e. quantities which do not scale with the size
of the system. As already pointed out, in a heavy-ion experiment all observables need to be constructed from the momenta and quantum numbers of the final state particles in each event. Furthermore, the number of produced particles typically increases with the centrality of the collision, i.e. with the volume of the system created. Therefore, the only way to devise volume independent observables is by considering ratios of observables which have the same volume (centrality) dependence. Examples which have been explored in experiments are ratios of particle numbers, mean energy or mean transverse momentum, i.e energy or transverse momentum divided by the number of particles. Thus, it will be useful to discuss the fluctuations of ratios in some more detail.

\subsection{Fluctuations of Ratios}
Consider the ratio of of two observables $A$ and 
$B$
\begin{eqnarray}
R_{AB} \equiv \frac{A}{B}
\end{eqnarray}
with\footnote{Defined this way, both $A$
and $B$ scale with $N$, the number of particles in the final state.} 
\begin{eqnarray}
A = \sum_{p,\alpha} a_p^\alpha n_p^\alpha 
\non
B = \sum_{p,\alpha} b_p^\alpha n_p^\alpha.
\end{eqnarray}
Expanding $A$ and $B$ around the mean,  $A = \ave{A} + \delta A$ and $B = \ave{B} + \delta B$, we get
\begin{eqnarray}
R_{AB}
=
\frac{\ave{A}}{ \ave{B}}
+
\frac{\ave{A}}{\ave{B}} \left( \frac{\delta A}{\ave{A}} - 
\frac{\delta B}{\ave{B}} \right)
+
O(\delta^2)
\label{eq:ratio_exp}
\end{eqnarray}
where $O(\delta^2)$ indicates terms that are of quadratic and higher orders
in $\delta A$ and $\delta B$.  Since $A$ and $B$ 
are extensive observables, the
neglected terms are at most ${\cal O}(1/\ave{N})$. From Eq.(\ref{eq:ratio_exp}), it is easy to see that to leading order in $\delta$
\begin{eqnarray}
\ave{R_{AB}} = \frac{\ave{A}}{\ave{B}}
\end{eqnarray}
and the variance
\begin{eqnarray}
\ave{(\delta R_{AB})^2} 
= \frac{\ave{A}^2}{\ave{B}^2} \lb 
\frac{ \ave{\delta A^2}}{\ave{A}^2} +  \frac{ \ave{\delta B^2}}{\ave{B}^2} 
- 2 \frac{\ave{\delta A \,\delta B}}{\ave{A}\ave{B}} \rb.
\label{eq:delta_r1}
\end{eqnarray}
Using the basic correlator $\Delta_{p,q}^{\alpha,\beta}$, 
this can be rewritten as
\begin{eqnarray}
\ave{(\delta R_{AB})^2}  = \frac{\ave{A}^2}{\ave{B}^2} 
\sum_{p,q} \sum_{\alpha,\beta}  \Delta_{p,q}^{\alpha,\beta}
\lb 
\frac{a_p^\alpha a_q^\beta}{\ave{A}^2} + 
\frac{b_p^\alpha b_q^\beta}{\ave{B}^2}  
-2 \frac{a_p^\alpha b_q^\beta}{\ave{A} \ave{B}} 
\rb.
\label{eq:ratio_fluct}
\end{eqnarray}
Often in the literature $\ave{\delta R_{AB}^2}$ is scaled by the square of the mean ratio to obtain a relative width, $\sigma^2_{A,B}$
\begin{eqnarray}
 \sigma^2_{A,B}\equiv\frac{\ave{(\delta R_{AB})^2}}{\ave{R_{AB}}^2}= \sum_{p,q} \sum_{\alpha,\beta}  \Delta_{p,q}^{\alpha,\beta}
\lb 
\frac{a_p^\alpha a_q^\beta}{\ave{A}^2} + 
\frac{b_p^\alpha b_q^\beta}{\ave{B}^2}  
-2 \frac{a_p^\alpha b_q^\beta}{\ave{A} \ave{B}} 
\rb
\label{eq:scaled_ratio_fluct}
\end{eqnarray}
which in case of an ideal gas simplifies to
\begin{eqnarray}
\sigma^2_{AB} (\rm ideal \, gas)
=  \sum_{p} \sum_{\alpha} \ave{n_p^\alpha} \omega_p^\alpha
\lb \frac{(a_p^\alpha)^2}{\ave{A}^2} + 
\frac{(b_p^\alpha)^2}{\ave{B}^2}  
-2 \frac{a_p^\alpha b_p^\alpha}{\ave{A} \ave{B}} \rb.
\label{eq:ratio_fluct_ideal}
\end{eqnarray}

In addition, covariances between two different ratio can be expressed in terms of the basic correlator in similar fashion
\begin{eqnarray}
\lefteqn{
\ave{\delta \lb \frac{A}{B} \rb  \delta \lb \frac{C}{D} \rb } =
\frac{\ave{A}\ave{C}}{\ave{B}\ave{D}} 
}
\non
&&
\sum_{p,q,\alpha,\beta} 
\lb 
\frac{a_p^\alpha c_q^\beta}{\ave{A}\ave{C}} 
+\frac{b_p^\alpha d_q^\beta}{\ave{B}\ave{D}}
-\frac{a_p^\alpha d_q^\beta}{\ave{A}\ave{D}} 
-\frac{b_p^\alpha c_q^\beta}{\ave{B}\ave{C}}
\rb
\Delta_{p,q}^{\alpha,\beta}
\label{eq:unlike_corr}
\end{eqnarray}
where
\begin{eqnarray}
C = \sum_{p,\alpha} c_p^\alpha n_p^\alpha, \,\,\,\,
D = \sum_{p,\alpha} d_p^\alpha n_p^\alpha.
\end{eqnarray}
This becomes useful, for example, in the discussion of charge dependent and charge independent transverse momentum fluctuations \cite{Adams:2003uw}.

The interesting physics, of course, resides in the dynamical correlations of the system. In order to expose these, one needs to compare to the purely statistical fluctuations of an uncorrelated system, which experimentally can be obtained via mixed event techniques. For ratio fluctuations, the variance for an uncorrelated system is given by the expression of the ideal gas, Eq.\ref{eq:ratio_fluct_ideal} with $\omega_p^\alpha=1$, and $\ave{n_p^\alpha}$ corresponding to the measured inclusive momentum distributions of the particles with quantum numbers $\alpha$. We can define the uncorrelated variance $\sigma^2_{\rm uncorrelated}$ as 
\begin{eqnarray}
 \sigma^2_{\rm uncorrelated}\equiv\sum_{p} \sum_{\alpha} \ave{n_p^\alpha} \omega_p^\alpha
\lb \frac{(a_p^\alpha)^2}{\ave{A}^2} + 
\frac{(b_p^\alpha)^2}{\ave{B}^2}  
-2 \frac{a_p^\alpha b_p^\alpha}{\ave{A} \ave{B}} \rb 
\end{eqnarray}
where here $\ave{\ldots}$ corresponds to actually measured (inclusive) averages. The difference between the actual variance of the system and that of an uncorrelated (mixed event) ensemble is usually referred to as $\sigma_{\rm dynamic}^2$ \cite{Voloshin:1999a} 
\begin{eqnarray}
\sigma_{\rm dynamic}^2 = \sigma^2 - \sigma^2_{\rm uncorrelated} .
\end{eqnarray}
The first such measure to be proposed has been the so called $\Phi$
variable \cite{Gazdzicki:1992ri,Mrowczynski:1998vt} which in terms of our variables
here is given by
\begin{eqnarray}
\Phi \equiv \ave{R_{AB}} \sqrt{\ave{N}}( \sigma  - \sigma_{\rm uncorrelated}).
\end{eqnarray}
The  ratio has also been proposed  \cite{Stephanov:1999zu} 
\begin{eqnarray}
F \equiv \frac{ \sigma^2}{\sigma^2_{\rm uncorrelated}}
\label{eq:F_ratio}
\end{eqnarray}
which has the advantage of being dimensionless and being less sensitive to acceptance effects, as we shall discuss in some more detail in the context of the fluctuations of the kaon-to-pion ratio below.
We finally note, that with these conventions, correlations due to quantum statistics already contribute to e.g. $\sigma_{\rm dynamic}$, typically at a few percent level, as discussed in \cite{Stephanov:1999zu,Mrowczynski:1998vt}.

\subsubsection{Examples: Fluctuations of the mean transverse momentum and particle ratios}
Event-by-event fluctuations of the mean transverse momentum, $p_t$, have been discussed in the literature as a measure of energy fluctuations, which should show a peak close to the QCD phase transition, where the specific heat has a maximum \cite{Stodolsky:1995ds,Shuryak:1998yj}. Transverse momentum fluctuations have also been discussed in the context of a search for the QCD critical point. As we will discuss in more detail in section \ref{sec:critical}, close to the critical point one expects long range fluctuations which would result in enhanced transverse momentum fluctuations, especially for small momenta \cite{Stephanov:1999zu,Stephanov:1998dy}. The signature in this case would be a maximum in the excitation function of $p_t$-fluctuations at the energy corresponding to the location of the critical point. Before we discuss some of the experimental results, let us first write down the necessary definitions. 
The mean transverse momentum is defined as
\be
p_t      &=& \frac{P_t}{N} = \frac{\sum_{p} n_p p_t(p)}{\sum_p n_p} 
\ee
where $(P_{t})$ denotes the transverse momentum of
an event with $N$ particles in the final state. Obviously, we are dealing with a ratio and with the substitution 
\begin{eqnarray}
A &=& P_t; \,\,\,a_p^\alpha=\sqrt{p_x^2+p_y^2}\equiv p_t(p)
\non
B &=& N;\,\,\,b_p^\beta=1
\end{eqnarray}
the above formalism, Eq.\ref{eq:scaled_ratio_fluct}, can be readily applied leading to the expression for the scaled variance 
\begin{eqnarray}
 \sigma^2_{p_t}=\sum_{p,q} \sum_{\alpha,\beta}  \Delta_{p,q}^{\alpha,\beta}
\lb 
\frac{p_t(p)\, p_t(q)}{\ave{P_t}^2} + 
\frac{1}{\ave{N}^2}  
-2 \frac{p_t(p)}{\ave{P_t} \ave{N}} 
\rb
\end{eqnarray}
which in the case of an ideal gas (see Eq.\ref{eq:ratio_fluct_ideal}) reduces to
\begin{eqnarray}
\sigma^2_{p_t}(\rm ideal\, gas)=\frac{1}{\ave{P_t}^2}\sum_{p,\alpha} \ave{n_p^\alpha} \omega_p 
\lb p_t(p) -\frac{\ave{P_t}}{\ave{N}} \rb^2.
\end{eqnarray}
We note that, since $\ave{P_t}$ scales with the number of particles in the final state, 
$\ave{P_t}\sim \ave{N}$, the variance scales as $\sigma^2_{p_t}\sim \frac{1}{\ave{N}}$, and thus is sensitive to the \emph{actual} acceptance of the detector. The same is true for all ratio fluctuations we discuss here.

\begin{figure}[ht]
\begin{center}
    \includegraphics[width=0.5\textwidth]{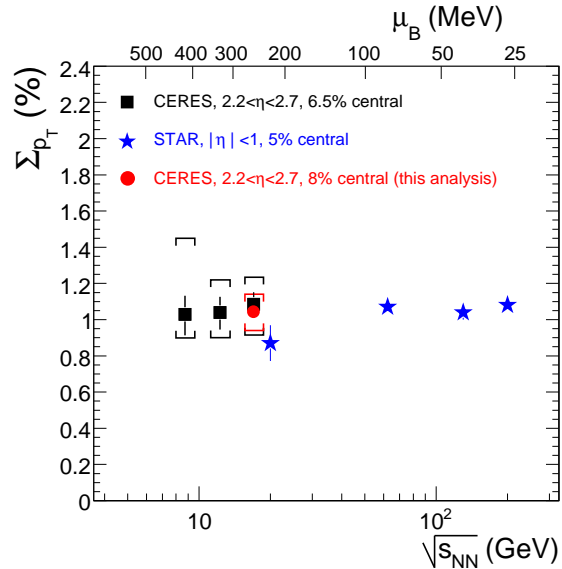}
\end{center}
\caption{Transverse momentum fluctuations for different center of mass energies.  Here $\Sigma_{p_T} = \sigma_{\rm dynamical}$ \cite{Adamova:2008sx}. The figure is adapted from \cite{Adamova:2008sx}.}
\label{fig:pt_fluct_data}
\end{figure}

Experimentally, $p_t$-fluctuations have been investigated by the CERES collaboration \cite{Sako:2004pw}, the NA49 collaboration \cite{Appelshauser:1999ft,Rybczynski:2008cv} at the CERN SPS and by the STAR \cite{Adams:2003uw}, and the PHENIX \cite{Adcox:2002pa,Adler:2003xq} collaborations at RHIC. These measurements cover a wide range of beam energies. The resulting $p_t$-fluctuations are  shown in Fig.\ref{fig:pt_fluct_data}. They are small  and show virtually no beam-energy dependence. 

In addition to the transverse momentum fluctuations for all charged particles,
one can investigate the $p_t$ fluctuations of the negative and positive charges
independently as well as the cross correlation between them. This aspect is discussed in detail in \cite{Jeon:2003gk} and has been investigated in experiment \cite{Adams:2003uw,Grebieszkow:2007xz}. In addition to the charge dependence the NA49 collaboration has also analyzed the $p_t$ fluctuations for different cuts in the transverse momentum (see Fig.\ref{fig:NA49_pt_fluct}) in order to test ideas about the QCD critical point, which we will discuss in section \ref{sec:critical}. 

\begin{figure}[ht]
\begin{center}
    \includegraphics[width=0.5\textwidth]{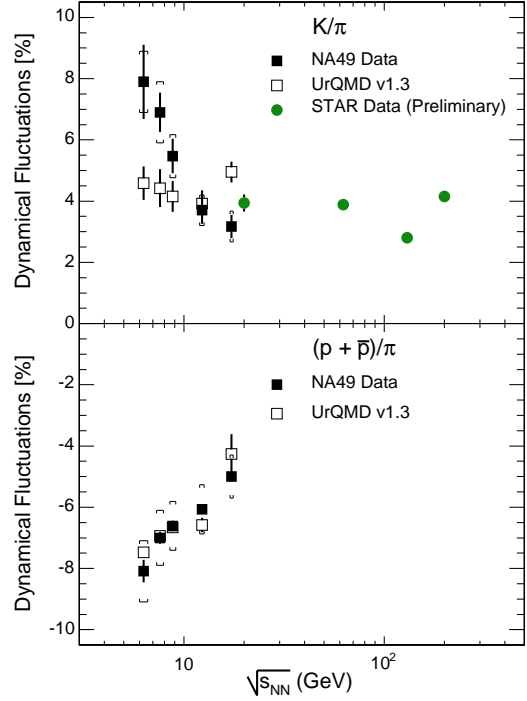}
\end{center}
\caption{Upper panel: Fluctuations of the $K/\pi$ ratio as a function of the center of mass energy. The data at low $\sqrt{s}$ are from the NA49 collaboration \cite{Alt:2008ca} and the ones at high $\sqrt{s}$ are prelimnary data from the STAR collaboration \cite{STAR_kpi}. Lower panel:
Fluctuations of the $p/\pi$ ratio as a function of the center of mass energy from NA49. Results from URQMD \cite{Bleicher:1999xi} calculations are shown as open squares. The figure is adapted from \cite{Alt:2008ca}.}
\label{fig:kpi_fluct}
\end{figure}

Let us next turn to the fluctuations of particle ratios and specifically to the fluctuations of the kaon-to-pion ratio, $K/\pi$. In this case
 \begin{eqnarray}
A &=& K; \,\,\,a_p^\alpha=\delta_{\alpha,K}
\non
B &=& \pi;\,\,\,b_p^\beta=\delta_{\beta,\pi}
\end{eqnarray}
so that
\begin{eqnarray}
 \sigma^2_{K/\pi}=\sum_{p,q} 
\lb 
	\frac{\Delta_{p,q}^{K,K}}{\ave{K}^2}
     +	\frac{\Delta_{p,q}^{\pi,\pi}}{\ave{\pi}^2}
     -2	\frac{\Delta_{p,q}^{K,\pi}}{\ave{K}\ave{\pi}}
\rb.
\end{eqnarray}
In the absence of any correlations, i.e a classical ideal gas, this reduces to
\begin{eqnarray}
 \sigma^2_{K/\pi}(\rm uncorrelated)=
\lb 
	\frac{1}{\ave{K}}
     +	\frac{1}{\ave{\pi}}
\rb.
\end{eqnarray}
The measurement of the $K/\pi$ fluctuations by the NA49 collaboration \cite{Afanasev:2000fu} were the first event-by-event fluctuations measurement in a heavy-ion experiment. The original motivation for this analysis was to look for separate event classes, for example, one with enhanced strangeness and one without. The observed fluctuations, however, are rather small, $\sim 2 \%$, indicating that the events generated in these collision are rather similar. Subsequently, the NA49 collaboration has measured the the $K/\pi$ fluctuations over the entire range of energies available at the CERN SPS \cite{Alt:2008ca}, and together with the preliminary measurements from STAR at RHIC \cite{STAR_kpi} we have an excitation function for this observable over a wide range of energies. This is depicted in the upper panel of Fig.\ref{fig:kpi_fluct}. We observe a rather steep increase of the $K/\pi$ fluctuations as the center of mass energy is decreased below $\sqrt{s}\sim 10 \, \rm AGeV$. This rise coincides with a maximum of the inclusive $K/\pi$ ratio \cite{Afanasiev:2002mx}. It can neither be explained in the statistical hadron gas model \cite{Torrieri:2007vv} nor can it be  reproduced by the URQMD model \cite{Bleicher:1999xi} (see Fig.\ref{fig:kpi_fluct}), which is able to describe the drop in the fluctuations of the proton-to-pion ratio, shown in the lower panel of Fig. \ref{fig:kpi_fluct}. 

Consequently, this rapid increase has sparked considerable discussion and speculations concerning the QCD critical point and the first order co-existence region, which we discuss in more detail in section \ref{sec:critical}. While this might indeed be a first experimental indication for interesting structures in the QCD phase diagram, we would like to raise a note of caution. As discussed in \cite{Jeon:1999gr}, ratio fluctuations scale roughly as the inverse of the \emph{accepted} multiplicity, 
\begin{eqnarray}
\sigma^2_{\rm dynamical} \sim \frac{1}{\ave{N}}_{accepted} .
\end{eqnarray}
Consequently, the observed rise may partially be due to the change of the actual acceptance with beam energy, which is always the case in a fixed target experiment such as NA49. As proposed in \cite{Jeon:1999gr}, the observed rise in the fluctuations would be more convincing if it showed up in the ratio of the measured variance over that from mixed event, as defined in Eq. \ref{eq:F_ratio}. This ratio does not exhibit any trivial multiplicity dependence. 

Finally let us briefly discuss the fluctuations of the ratio of positively over negatively charged particles, which is directly related to the net-charge fluctuations discussed in section \ref{sec:conserved}. These fluctuations are a direct probe for the existence of a QGP as pointed out in \cite{Jeon:2000wg,Asakawa:2000wh} and the many subtleties such as global charge conservation \cite{Koch:2001zn} have been discussed in the literature (for a review see \cite{Jeon:2003gk}). 
The fluctuations of the ratio of positively over negatively charged particles 
\begin{eqnarray}
R_{+-} &=& \frac{N_+}{N_-}
\end{eqnarray}
is given by
\begin{eqnarray}
   \sigma^2_{+-}=\lb 
\frac{\ave{\delta N_+^2}}{\ave{N_+}^2} + \frac{\ave{\delta N_-^2}}{\ave{N_-}^2} 
-2 \frac{\ave{\delta N_+ \, \delta N_- }}{\ave{N_+} \ave{N_-}}
\rb
\end{eqnarray}
which in the limit of small net charge 
\begin{eqnarray}
\ave{Q}=\ave{N_+ - N_-}\ll\ave{N_{ch}}=\ave{N_+ + N_-}
\end{eqnarray}
can be written as
\begin{eqnarray}
\sigma^2_{+-}=\frac{4 }{\ave{N_{ch}}^2} \ave{\delta N_+^2 + \delta N_-^2 - 2\delta N_+ \delta N_-}
=\frac{4 \ave{\delta Q^2}}{\ave{N_{ch}}^2} .
\end{eqnarray}
Since the number of charged particles is a measure for the entropy, $\ave{N_{ch}}\sim S$, the observable $D \equiv \ave{N_{ch}}\sigma^2_{+-}$ measures the charge fluctuation per degree of freedom, which, as discussed in section \ref{sec:conserved}, changes by roughly a factor of three when going from a hadron gas to the a Quark Gluon Plasma. As already mentioned, for this observable to be useful, a clear separation of scales in rapidity is necessary, otherwise hadronization and subsequent hadronic interactions degrade the signal. This observable has been measured in experiment both at the SPS and at RHIC \cite{Alt:2004ir,Appelshauser:2004ms,Adcox:2002mm,Abelev:2008jg} by the CERES, NA49, PHENIX, and STAR collaborations. While the RHIC results are consistent with the predictions for a resonance gas, the results at the SPS are difficult to interpret as the necessary separation of rapidity scales at these energies is not satisfied and global charge conservation effects dominate the signal. No indication of a \QGP has been seen, however.

\section{Fluctuations and the QCD critical point}
\label{sec:critical}
As pointed out in the introduction, fluctuations and correlations are unique signatures for phase transitions. Therefore, an experimental search for a possible critical point and a first order co-existence region in the QCD phase diagram is intimately connected with the study and measurement of fluctuations and correlations. In this section we will discuss how presently available data constrain the existence of a critical point and what additional measurements should be done  to shed more light on this very interesting question. 

Let us start with a brief reminder of the physics and the present theoretical understanding of the QCD critical point. A detailed account can be found in \cite{Stephanov:2004wx}. The basic argument which leads to the existence of a critical point in the QCD phase diagram is rather straightforward. For vanishing baryon-number chemical potential Lattice QCD calculations have established that the transition is an analytic cross-over \cite{Aoki:2006we}. If the transition at zero temperature but finite density is of first order, as most models predict, then the first order phase transition line in the $T-\mu_{\rm Baryon}$ plane has to end. This ``end-point'' is the QCD critical point associated with a second order transition. This scenario is depicted in Fig. \ref{fig:phase_diagram}. The question of course arises how robust this simple argument really is. The cross-over transition at $\mu_B=0$ seems well established by Lattice QCD. The first order transition at $T=0$, on the other hand, is model dependent. While essentially all chiral models do predict such a first order transition at $T=0$ and finite $\mu_{\rm baryon}$ one should keep in mind that all these models\footnote{For a compilation of the model predictions see \cite{Stephanov:2004wx}.} are essentially  extensions of the linear sigma model, such as the Nambu model with or without Polyakov loop dynamics, and small modifications may already alter the conclusions. For example, once the t'Hooft anomaly interaction is taken into account, the first order transition at $T=0$ may turn into a cross-over with yet another critical point at low temperatures and high density \cite{Hatsuda:2006ps}. 

\begin{figure}[ht]
\begin{center}
        \includegraphics[width=0.45\textwidth]{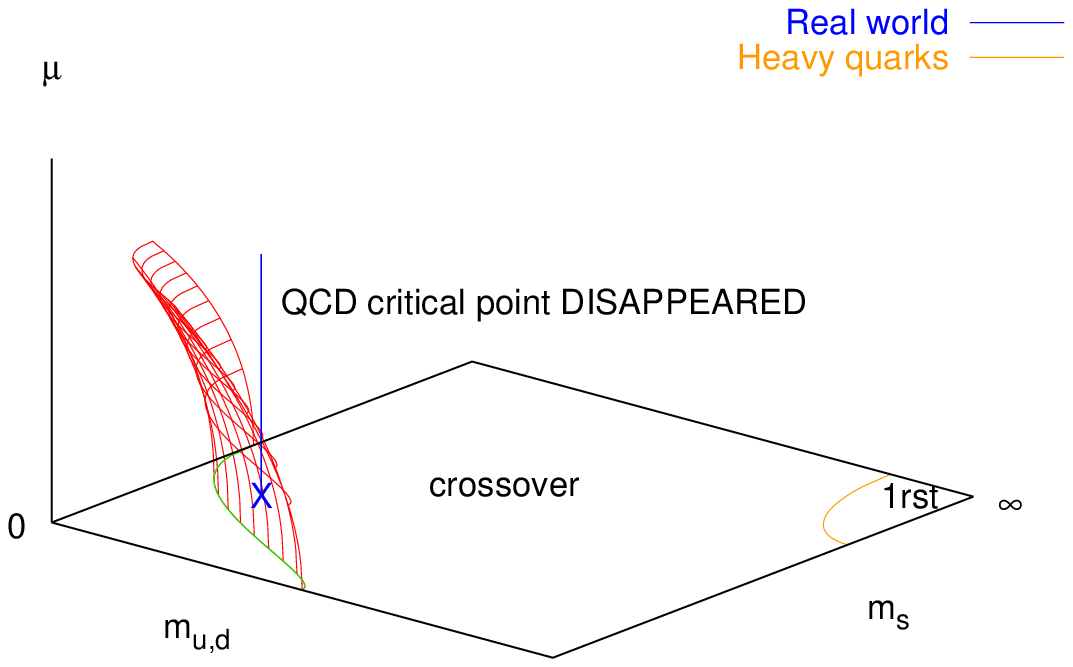}
	\includegraphics[width=0.45\textwidth]{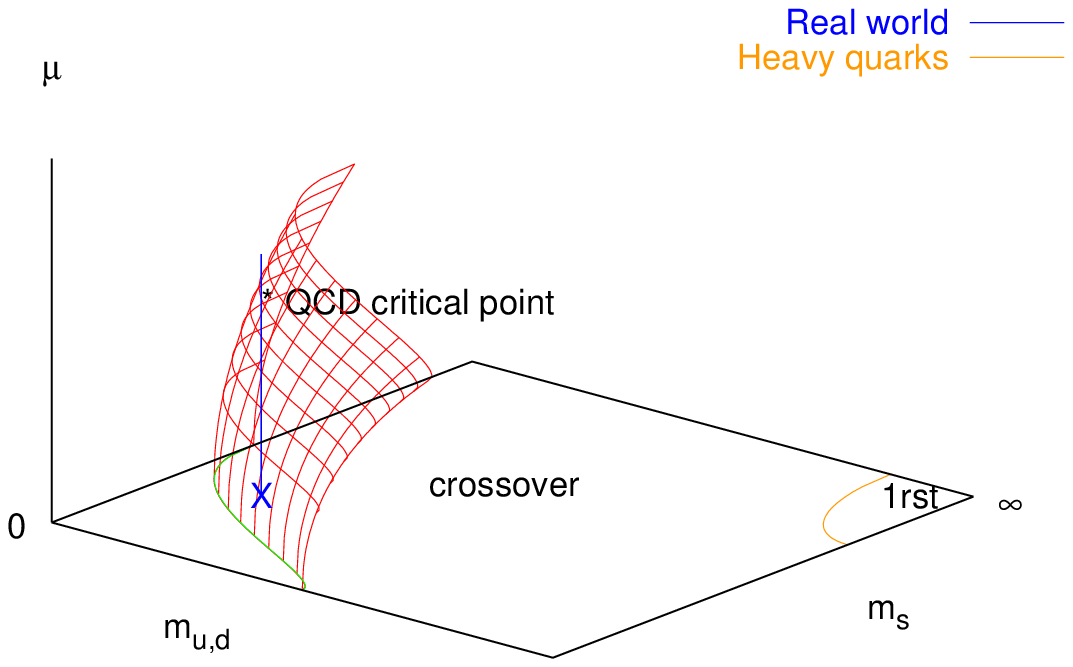}
\end{center}
\caption{Behavior of the critical line as a function of chemical potential $\mu_B$. Left panel: Scenario favored by Lattice QCD \cite{deForcrand:2006pv,deForcrand:2008vr} where critical line moves towards smaller quark masses. Right panel: Standard scenario predicted by most chiral models, where critical line moves towards higher quark masses. The figure is adapted from \cite{deForcrand:2006pv}.}
\label{fig:forcrand}
\end{figure}

A nice way to analyze these models and also Lattice QCD results in the region of small chemical potentials is to find the critical quark mass for which one obtains a second order transition \cite{deForcrand:2006pv}. This is depicted in Fig.\ref{fig:forcrand}. Most chiral models predict that the critical quark mass increases with the chemical potential (right panel of Fig.\ref{fig:forcrand}). In this case, one expects a critical point at finite chemical potential once the critical quark mass coincides with the physical quark mass, as can be seen in the figure. Lattice QCD, on the other hand, seems to favor the opposite trend, namely a decreasing critical quark mass (left panel of Fig.\ref{fig:forcrand}). This is the result of \cite{deForcrand:2006pv,deForcrand:2008vr} obtained in an expansion up to fourth order in the chemical potential $\mu_B$ on a  rather small lattice. If these lattice results hold up for larger lattices it seems that the chiral dynamics does not predict the small $\mu_B$ behavior of the critical quark mass correctly and other effects are more dominant. One possibility would be a repulsive vector coupling, which is neither constrained nor ruled out by symmetry arguments. As shown in \cite{fukushima08} a suitable choice of a repulsive vector coupling can indeed reproduce the trend seen on the lattice.

There have also been attempts to find the critical point directly in Lattice QCD. This requires reweighting techniques in order to circumvent the problems associated with the complex fermion determinant at finite chemical potential. The pioneering work \cite{Fodor:2001pe} in this approach has indeed located a critical point. With realistic quark masses, this method predicts its location at $T \simeq 160 MeV$ and $\mu_B \simeq 360 \, \rm MeV$ \cite{Fodor:2004nz}. However, the method employed can not easily be extended to larger volumes and, therefore, one does not know if the signal survives in the infinite volume limit. Other approaches calculate the free energy at finite chemical potential as a Taylor expansion in terms of the chemical potential (see e.g \cite{Allton:2005gk}, \cite{Gavai:2008zr}). As discussed in section \ref{sec:conserved} the expansion coefficients are given by the baryon number cumulants or susceptibilities. While this method does not allow to extract a critical point directly it can provide limits for its location. At present a conservative limit for its chemical potential is $\mu_B\gtrsim 2 T_c$ (\cite{Karsch:2008fe,Gavai:2008zr}), where $T_c\sim 180 \, \rm MeV$  is the transition temperature at vanishing baryon number density. 

Obviously there is only limited theoretical guidance for an experimental search for the critical point as the model predictions for its location vary quite a bit. From hadronic freeze out systematics \cite{Braun-Munzinger:2003zd}, on the other hand, one knows that the chemical potential of the system created depends on the center-of-mass energy of the collision.  Unless the temperature of the critical point is unexpectedly low, one can explore regions up to about $\mu_B \leq 500 \, \rm MeV$ in the chemical potential by lowering the beam energy to about $\sqrt{s}\simeq 5 \,\rm GeV$. Hence, the strategy for a search is  to study excitation functions of various observables and see if they show non-monotonic behavior at the same beam energy, indicating the location of the critical point or of the first order phase co-existence region 

Since the baryon density is  an order parameter for the phase transition at finite density, baryon number fluctuations are the natural observable to consider. In  section \ref{sec:conserved} we have already discussed the serious limitations baryon number conservation imposes on this observable, especially for low center-of-mass energies. In addition, the measurement of the baryon number requires the detection of neutrons, which is difficult. However, as argued in \cite{Hatta:2003wn}, it may be sufficient to study proton number fluctuations, as the iso-vector channel does not show critical behavior. This may also soften the limitations due to global baryon-number conservation. 
However, even if the system reaches the critical point, the correlation length, which diverges in a thermal system, would be finite due to critical slowing down together with the finite time the system has to develop the correlations. In \cite{Berdnikov:1999ph} a correlation length of $\xi \simeq 2.5 \,\rm fm$ has been estimated based on these considerations.  Therefore, it may be advantageous to study higher order cumulants which depend on higher powers of the correlations length \cite{stephanov08}. Indeed the fourth order cumulant  $\chi^{(4)}$, Eq. \ref{eq:chi_4}, scales like the seventh power of the correlation length, $\sim \xi^7$ \cite{stephanov08}. Thus if the correlation length increases only by 10\%  in the vicinity of the critical point, one should see an enhancement by a factor of two in the fourth order cumulant, whereas the second order cumulant, i.e. the fluctuations, would only increase by 20\%.

\begin{figure}[ht]
\begin{center}
    \includegraphics[width=0.9\textwidth]{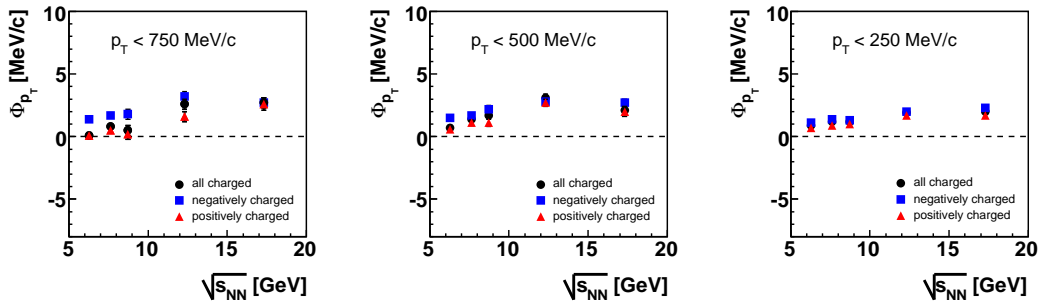}
\end{center}
\caption{Preliminary data on the energy dependence of $p_t$ fluctuations from the NA49 collaboration \cite{Grebieszkow:2007xz} for all charged particles and for positively and negatively charged particles. The panels show the fluctuations for different cuts in the transverse momentum. The figure is adapted from \cite{Grebieszkow:2007xz}.}
\label{fig:NA49_pt_fluct}
\end{figure}

Initially transverse momentum fluctuations have been proposed as a signature for the critical point \cite{Stephanov:1998dy,Stephanov:1999zu}, since close to the critical point the system should develop large, and mostly long range, i.e low momentum, fluctuations. Therefore, it was suggested that an excitation function of the transverse momentum fluctuations should show non-monotonic behavior, especially for small transverse momenta. In the meantime, such an excitation function has been measured and it is shown in Fig.\ref{fig:NA49_pt_fluct} for different charge combinations and different cuts on the transverse momentum. Critical fluctuations, corrected for critical slowing down and expansion of the system \cite{Berdnikov:1999ph} would lead to a bump which should be at least a factor of two larger than the statistical background. Obviously, the data shown in Fig.\ref{fig:NA49_pt_fluct} do not show such a behavior, even for small transverse momenta. The results at RHIC \cite{Adams:2003uw}, shown in Fig.\ref{fig:pt_fluct_data}, are consistent with the data from SPS. Hence, so far there is no indication of a critical point in the transverse momentum fluctuation measurements. Of course it could be that the signal is too weak to be seen and it may also be washed out by subsequent hadronic interactions. To address this issue, higher cumulants, as discussed above, need to be measured as they should show a stronger enhancement close to the critical point. 
Furthermore, on the theoretical side, one needs to get a better understanding of the degradation of the proposed signals in the hadronic phase.

The only observable which shows a strong beam energy dependence are the fluctuations of the kaon-to-pion ratio, shown in Fig.\ref{fig:kpi_fluct}. As discussed in section \ref{sec:observable}, neither a transport approach nor the statistical hadron gas model can reproduce these data. However, as already pointed out, $\sigma_{\rm dynamic}^2$ scales with the inverse \emph{accepted} multiplicity, and the observed rise may very well be partially due to the changing acceptance of the fixed target NA49 experiment. If the enhancement turns out to be real, it is still not clear if is related to the critical point. Close to the critical point one would expect the fluctuations of the pion number to be enhanced. But this would imply also enhanced fluctuations of the proton-to-pion ratio, which is not observed in experiment, as shown in the lower panel of Fig.\ref{fig:kpi_fluct}. There the data are well reproduced by the URQMD calculations.

Although this section is mostly about the critical point and its detection, let us emphasize that it might be more beneficial to look for and identify the first order co-existence region. Contrary to the critical {\em point}, the first order transition corresponds to an entire {\em region} in the $T-\rho$ phase diagram. Thus it is more likely for the system to cross this region rather than the critical point. It is also more likely for the system to spend sufficient time in this region in order to develop measurable effects. One example is  the development of spinodal instabilities, which are a generic phenomenon of dynamical first order transitions \cite{Chomaz:2003dz}. Spinodal instabilities have been studied and successfully identified in the context of the nuclear liquid gas phase transition \cite{Borderie:2001jg}. In the case of the QCD first order transition, spinodal instabilities could lead to kinematic correlations among particles \cite{Randrup:2005sx} and to enhanced fluctuations of strangeness \cite{Koch:2005pk}. And indeed the observed enhancement of the kaon-to-pion fluctuations, if real, may be due to these enhanced fluctuations in the strangeness sector \cite{Koch:2005pk}. However, just as for the critical point, there has been no quantitative calculation of the effect due to hadronic re-scattering on the observables. In addition, so far there is no dynamical model which carries the system through the spinodal region, as it is the case for the nuclear liquid gas phase-transition, where such as model proved to be extremely useful in guiding the experimental searches. These models have also helped to develop unique observables, such as the variance of the cluster size which subsequently could be identified in experiment\cite{Borderie:2001jg}. This analysis lead to a rather convicing case for the existence of a first order phase co-existence region for nuclear matter. 

Although this article is about fluctuations, in closing let us briefly mention other observables which are discussed in the context of the critical point. Most prominently is the idea to look for soft modes in the low-mass dilepton invariant mass spectrum. However, it is not clear if the soft modes, responsible for the large density fluctuations close to the critical point, are visible in the dilepton channel, since they are of space like origin. Indeed, an analysis of the fluctuations close to the critical point carried out in the Nambu model with finite quark masses \cite{Fujii:2004jz,Fujii:2004za} shows that the sigma-meson remains gaped at the critical point, contrary to the chiral transition in the limit of vanishing quark masses. Thus, in this model no significant change due to the critical point has been seen in the time-like spectrum, which is accessible to dilepton spectroscopy.

There are a number of other possibilities which have not yet been explored theoretically. For example, it maybe interesting to further explore the co-variances between the baryon density fluctuations and other quantities which couple to the baryon density, such as e.g. dileptons. These co-variances are expected to become large and could possibly be developed into practical observables, which will not be affected by baryon number conservation.

\section{Summary and Conclusions}
\label{sec:conclude} 
In this article we have provided a review on some selected aspects concerning the study of correlations and fluctuations in heavy ion collisions. We have discussed some of the recent lattice results on fluctuations and correlations. While the results are still improving, a rather interesting picture for the matter above the transition temperature emerges from these Lattice calculations. The ratio of flavor off-diagonal to diagonal susceptibilities as well as the fourth order cumulants in baryon number and charge indicate that the quark degrees of freedom behave like uncorrelated particles at temperatures above $T \gtrsim 1.3 \,T_c$. Furthermore, the transition from a hadron gas seem rather rapid. To which extent there are some additional, non-trivial features just below the transition temperature is, at present, an open question. While the most recent $N_t=6$ data with almost physical quark masses show a rather featureless transition from hadron to independent quarks, the results for a coarser, $N_t=4$, lattice exhibit a peak in the fourth order cumulant for the baryon number. This, if correct, would be a very nice first indication for a critical point at finite chemical potential. If, on the other hand, this peak disappears in the continuum limit, as the $N_t=6$ data seem to suggest, the critical point will likely be located at much higher chemical potential. We note, that the featureless, rapid transition from hadrons to quarks, would nicely explain the observed quark number scaling and recombination phenomenology observed at RHIC.

As far as observables are concerned we have discussed electric charge fluctuations, transverse momentum fluctuations, and the fluctuations of the kaon-to-pion ratio. All these have been measured over a wide range of beam energies, from the CERN SPS to RHIC, and none of the excitation functions, with the possible exception of the kaon-to-pion ratio, show any significant beam energy dependence. In the case of the kaon-to-pion ratio a rapid rise towards the lowest energies is seen, which may or may not be due to simple scaling of the observable with the acceptance. If the observed rise is real, then this may very well be the first hint for some non-trivial phase structure probed at these lowest energies.

Concerning the QCD critical point, the present data set does not provide any evidence for its existence in the region probed so far. The transverse momentum fluctuations do not show any non-monotonic behavior, as originally predicted. However, it could very well be that the signal is too weak, as there is not sufficient time to develop a large correlation length in these systems. Therefore, it is imperative to measure the higher cumulants as well. For example, the fourth order cumulant scales like the seventh power of the correlation length, whereas the second order, which controls the transverse momentum fluctuations, only scales like the square of the correlation length. Fortunately, the RHIC accelerator can easily cover the range of energies required for a comprehensive exploration of the QCD phase diagram. And, with the two state of the art detectors, PHENIX and STAR, a definitive measurement of these and other observables is possible. In addition the CBM experiment at the FAIR facility will be able to perform high statistics measurements in the region of high chemical potential, allowing for a detailed spectroscopy of this most interesting region of the QCD phase diagram.
 
Let us conclude by pointing out the obvious: A dedicated energy scan program at RHIC and FAIR would be very desireable to definitively answer the central question of the field: Is there a non-trivial phases structure of QCD?

\section{Acknowledgements}
The writing of this review has benefitted from the many presentations and discussion at the INT program entitled ``The QCD critical point''. The author thanks L. Ferroni for a critical reading of the manuscript. This work was supported  by the Director, Office of Energy Research, Office of High Energy and Nuclear Physics, Divisions of Nuclear Physics, of the U.S. Department of Energy under Contract No. DE-AC02-05CH11231.

\providecommand{\href}[2]{#2}\begingroup\raggedright\endgroup

\end{document}